\documentclass[aps,pra,reprint]{revtex4-2}
\usepackage[utf8]{inputenc}
\usepackage{graphicx}
\usepackage{amsmath}
\usepackage{comment}
\usepackage{siunitx}
\usepackage[caption=false]{subfig}
\usepackage{subfiles}
\usepackage{svg}
\usepackage{quantikz}
\usepackage{bm}%

\usepackage[colorlinks]{hyperref}
\hypersetup{%
        plainpages=true,
        breaklinks=true,
        hypertexnames=false,
        pageanchor=true,
        colorlinks=true,
        linkcolor={blue},
        citecolor={blue},
        urlcolor={blue},
        anchorcolor={black}
      }

\renewcommand{\eqref}[1]{\mbox{Eq.~(\ref{#1})}}
\newcommand{\figref}[1]{\mbox{Fig.~\ref{#1}}}
\newcommand{\figpanel}[2]{Fig.~\hyperref[#1]{\ref*{#1}(#2)}} 
\newcommand{\figpanels}[3]{Fig.~\hyperref[#1]{\ref*{#1}(#2)-(#3)}} 
\newcommand{\figpanelNoPrefix}[2]{\hyperref[#1]{\ref*{#1}(#2)}} 
\newcommand{\figpanelsNoPrefix}[3]{\hyperref[#1]{\ref*{#1}(#2)-(#3)}} 

\usepackage{mleftright}

\begin{document}
\title{Heralding entangled optical photons from a microwave quantum processor}
\author{Trond Hjerpekjøn Haug}
\email{trond.haug@chalmers.se}
\author{Anton Frisk Kockum}
\author{Rapha\"{e}l Van Laer}
\email{raphael.van.laer@chalmers.se}
\affiliation{Department of Microtechnology and Nanoscience, Chalmers University of Technology, 412 96 Gothenburg, Sweden}
\date{\today}

\begin{abstract}
Exploiting the strengths of different quantum hardware components may enhance the capabilities of emerging quantum processors. Here, we propose and analyze a quantum architecture that leverages the non-local connectivity of optics, along with the exquisite quantum control offered by superconducting microwave circuits, to produce entangled optical resource states. Contrary to previous proposals on optically distributing entanglement between superconducting microwave processors, we use squeezing between microwaves and optics to produce microwave-optical Bell pairs in a dual-rail encoding from a single microwave quantum processor. Moreover, the microwave quantum processor allows us to deterministically entangle microwave-optical Bell pairs into larger cluster states, from which entangled optical photons can be extracted through microwave measurements. Our scheme paves the way for small microwave quantum processors to create heralded entangled optical resource states for optical quantum computation, communication, and sensing using imperfect microwave-optics transducers. We expect that improved isolation of the superconducting processor from stray optical fields will allow the scheme to be demonstrated using currently available hardware.
\end{abstract}

\maketitle

\section{Introduction}  

Entanglement is a central resource for quantum computation, communication, and sensing~\cite{Horodecki2009}. For quantum technology platforms such as superconducting microwave circuits~\cite{blais_circuit_2021, Gu2017}, continuous-variable optics~\cite{asavanant_generation_2019, larsen_deterministic_2019}, and trapped ions~\cite{leibfried_quantum_2003, benhelm_towards_2008, bruzewicz_trapped-ion_2019}, controlled generation of entanglement is now a routine task, while for discrete-variable optics it remains a daunting challenge~\cite{kok_linear_2007}. Optical entanglement can be produced probabilistically through fusion measurements~\cite{browne_resource-efficient_2005,varnava_how_2008,bartolucci_fusion-based_2023} on single optical photons heralded from second-~\cite{zhao_high_2020, zhang_high-performance_2021, cabrejo-ponce_ghz-pulsed_2022} and third-order~\cite{Ma_17,paesani_near-ideal_2020, steiner_ultrabright_2021} spontaneous biphoton emission.
Alternatively, optical quantum emitters can entangle optical photons deterministically via spin-photon interactions~\cite{lindner_proposal_2009, economou_optically_2010, schwartz_deterministic_2016, uppu_quantum-dot-based_2021, cogan_deterministic_2023, cao_photonic_2024,chen_heralded_2024}.
A less explored path to produce entanglement between single optical photons is to prepare the entangled state in a mature platform with a strong quantum nonlinearity---such as superconducting microwave circuits---and then transduce the state to optics. This approach places strict performance requirements on microwave-optics transducers~\cite{wang_using_2012, zhong_microwave_2022}. Despite rapid progress in recent years, current transduction platforms cannot yet directly convert quantum states between microwaves and optics~\cite{jiang_efficient_2020,mirhosseini_superconducting_2020,jiang_optically_2023,weaver_integrated_2023,sahu_entangling_2023, meesala_non-classical_2024}. 

In this work, we present a scheme for producing entangled optical states, called resource states, using microwave-optics transducers. Each qubit in the resource state is encoded as a dual-rail erasure qubit---the preferred code for discrete-variable optical quantum technologies. Our scheme goes beyond previous proposals for heralding single microwave-optical photon pairs~\cite{zhong_proposal_2020, zhong_entanglement_2020}: we propose a full architecture for a microwave quantum processor to produce multi-qubit optical resource states. The architecture is designed to meet the requirements set by fault-tolerant optical quantum technologies such as all-optical quantum repeaters~\cite{azuma_all-photonic_2015} and fusion-based quantum computers~\cite{bartolucci_fusion-based_2023}. 
\begin{figure*}
    \centering
    \includegraphics[width=\linewidth]{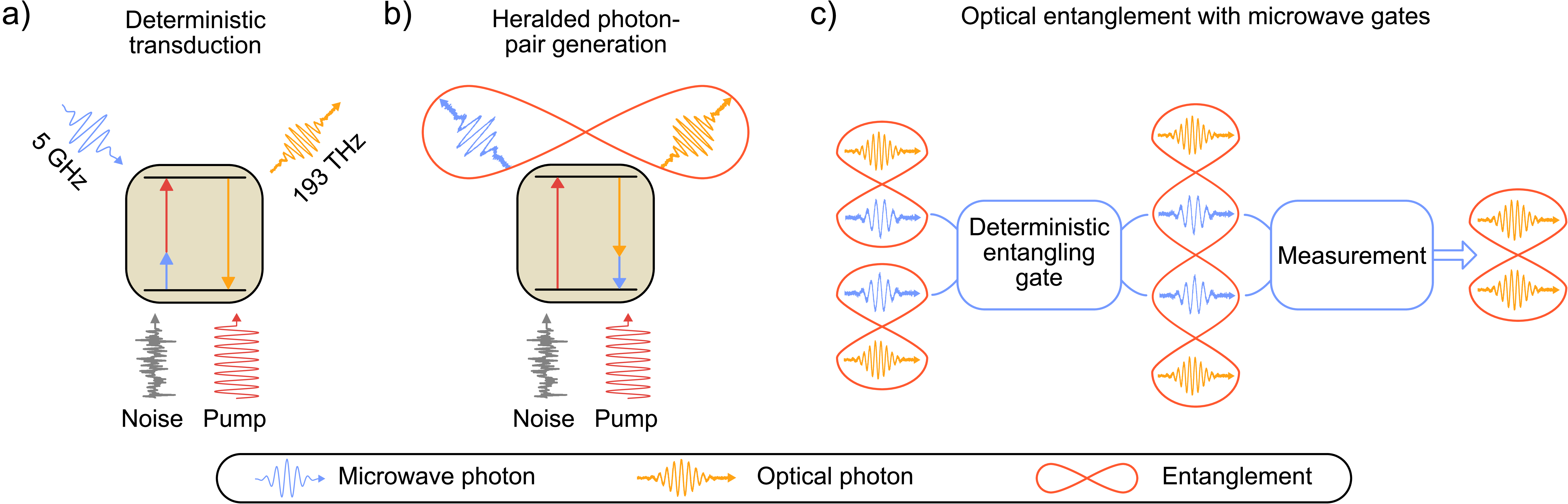}
    \caption{\textbf{Generating optical entanglement through microwave-optics photon pair generation and deterministic microwave gates.} (a) High-performance microwave-optics transducers could convert incoming microwave photons to optical photons using an optical pump to bridge the energy gap. Trade-offs between added noise and efficiency prevent today's transducers from converting quantum states deterministically. (b) Even imperfect microwave-optics transducers can be operated as high-fidelity probabilistic photon-pair generators. In this work, the presence of a microwave photon --- detected through a microwave parity check --- heralds an entangled microwave-optical photon pair. (c) We propose to leverage deterministic microwave gates on several such pairs to generate optical entanglement. The entanglement structure of the full microwave-optical quantum state can be transferred to an all-optical quantum state by measurements in the microwave domain. This allows deterministic entanglement swapping from microwave-optical to all-optical states.}
    \label{fig:transduction-VS-SPDC}
\end{figure*}
We present the fundamental building block in the architecture, which heralds microwave-optical Bell pairs with spectrally pure optical photons. We show how such microwave-optical Bell pairs can be naturally interpreted as the vertices of microwave-optical graph states constructed using the superconducting processor, and how measurements of the superconducting qubits produce an all-optical graph state identical to the hybrid microwave-optical graph state. Finally, we find a set of performance parameters for each component in our architecture where all-optical resource states can be efficiently constructed with error rates near the threshold for fault-tolerant fusion-based quantum computation. We describe in detail how hardware imperfections translate into qubit errors that must be handled by the quantum error-correcting code, and we identify optical loss and thermal noise in the transducer as the main areas of improvement for future experimental implementations. 

The presented scheme enables a modular, fault-tolerant quantum computer with superconducting processors as resource-state generators (RSGs) seeding an optical processor that implements logic and error correction~\cite{bombin_interleaving_2021}. It takes advantage of the recent development of dual-rail architectures for superconducting quantum processors that aims to make such processors capable of fault-tolerant quantum computing~\cite{teoh_dual-rail_2023, kubica_erasure_2023, Levine2024}. However, the construction of a single RSG is much simpler than that of a fault-tolerant superconducting processor. RSGs capable of producing resource states of $n$ optical dual-rail qubits require only $\sim n/p$ physical superconducting dual-rail qubits, where $p$ is the microwave-optical heralding probability per clock cycle. With the proposed architecture, we simulate the preparation of a microwave-optical Bell state and find that $p \lesssim \SI{20}{\percent}$ for practical applications. Therefore, integration of $\sim 100$ dual-rail qubits on a single chip would be sufficient to build RSGs capable of producing optical resource states with $n\sim 20$ qubits for fault-tolerant fusion-based quantum computation~\cite{ bell_optimizing_2023, bombin_fault-tolerant_2023}. This proposal provides optical quantum processors with the main element lacking in their architecture: a fast source of low-overhead optical entanglement. It is, to the best of our knowledge, the first to propose and study microwave quantum processors as resource-state generators, i.e., ``entanglement factories'' for optical quantum technologies. The well-known schemes for connecting microwave qubits through optically mediated entanglement~\cite{krastanov_optically_2021,duan_long-distance_2001} use only limited optical technology. This work initiates a wider research landscape where quantum processing can take place in both the optical and microwave domain.

This modular approach to building a fault-tolerant quantum processor has several potential advantages compared to building a full-size processor exclusively from superconducting qubits. First, a single RSG needs fewer physical qubits than fault-tolerant microwave processors, thus bypassing the need to scale such processors beyond the current state of the art. Second, no logical information is stored on the microwave processor which can be corrupted by ionizing radiation or high-energy impacts~\cite{wilen_correlated_2021, mcewen_resolving_2022}. Such events would temporarily halt the production of resource states from an RSG, but this could be compensated by routing resource states from a different RSG with minimal overhead. Third, the traveling optical photons in the resource states can be used to implement high-threshold surface codes concatenated with efficient low-density parity-check (LDPC) codes that require non-local connections~\cite{breuckmann_quantum_2021, bombin_logical_2023}. We can make further use of the non-local connections to implement an active-volume logical architecture~\cite{litinski_active_2022}. This reduces the resource cost for a quantum computation by minimizing the number of idling qubits during a computation, with potentially orders-of-magnitude gains for large computations. Fourth, the logical processing is done entirely through destructive measurements, which implies that qubit leakage (e.g., transmons excited to the state $\vert f\rangle$) does not cause errors that stabilizer fault tolerance is not designed to handle~\cite{fowler_coping_2013, varbanov_leakage_2020}.

In the near-term, the use of imperfect microwave-optics transducers requires a mixed heralded-deterministic scheme (\figref{fig:transduction-VS-SPDC}) to generate the optical entanglement from heralded microwave-optical qubit pairs and deterministic microwave gates. If microwave-optics transducers continue to improve on their current steep path \cite{jiang_efficient_2020,mirhosseini_superconducting_2020,delaney_superconducting-qubit_2022,jiang_optically_2023,weaver_integrated_2023,sahu_entangling_2023,meesala_non-classical_2024}, the optical entanglement can eventually be generated deterministically through direct transduction of the entangled microwave photons. In contrast to purely microwave-frequency approaches to exploit LDPC codes \cite{bravyi_high-threshold_2024} and hardware-efficient resource states using tailored delay lines \cite{ferreira_deterministic_2024}, optical photons are ideal for the non-local connectivity needed in LDPC codes given their low noise, loss, and crosstalk in widely available optical fiber delay lines even at room temperature. This proposal takes a step forward on the overarching question: How can the strengths of optical and microwave photons as quantum information carriers be combined and harnessed optimally?

\begin{figure*}
    \centering
    \includegraphics[width=\linewidth]{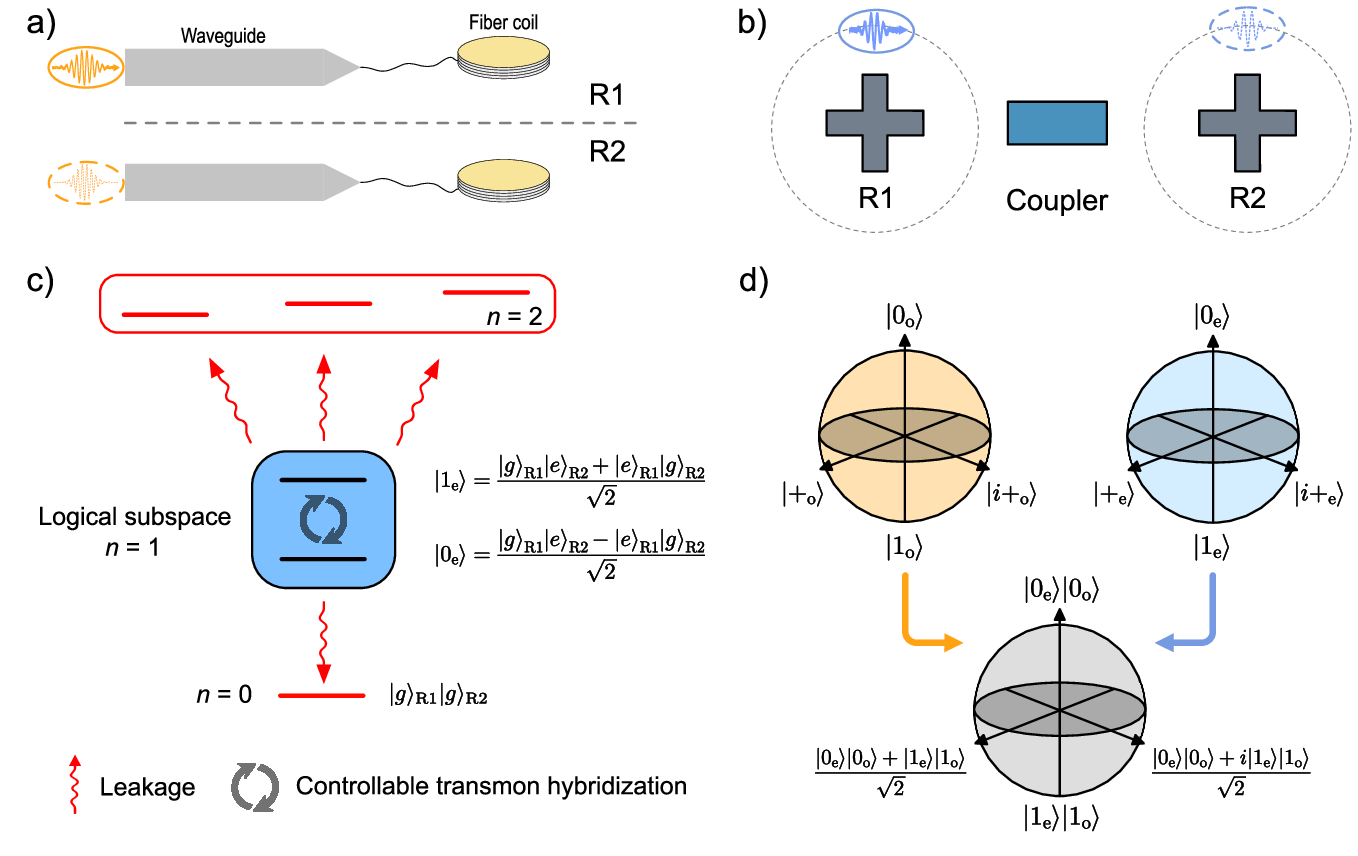}
    \caption{\textbf{A dual-rail qubit implemented with optics and transmons.} (a) An optical dual-rail qubit consists of a single optical photon in one of two modes, here represented with two physically separated optical waveguides. The waveguides can be interfaced with optical fiber where photons can be stored before they are measured. We use the label R1 the first rail of the dual-rail qubit, and R2 for the second rail. (b) Example implementation of a microwave dual-rail with transmon qubits (crosses) and a coupling element to control the interaction between them. A single microwave photon is stored in the combined system of the two transmons. When we measure the state of each transmon separately, we will find the photon either in R1 or R2. (c) Energy diagram of the dual-rail qubit in (b), with the dual-rail logical subspace highlighted in dark blue and leakage states in red. The number of microwave photons is indicated by $n$ for each state. Two hybridized transmons form a two-level system with a controllable gap between the symmetric and antisymmetric single-excitation states, which we use to define the computational basis states of the dual-rail qubit. Photon annihilation and creation bring the system outside the logical subspace, which can be detected with a parity measurement. (d) Bloch-sphere representation of the logical subspace of the dual-rail qubit. The optical and microwave dual-rail qubits can be combined to form a single, hybrid qubit. All hybrid-qubit states apart from the Pauli-\textit{Z} eigenstates are entangled. The hybrid-qubit picture of entangled microwave and optical qubits is useful when constructing resource states, as explained in Section~\ref{sec:protocol}.}
    \label{fig:Bloch}
\end{figure*}

\section{Hardware implementation}

Our proposal is based on a dual-rail encoding of qubits in both the microwave and optical domains. On the optical side, the dual-rail qubit is a flying optical photon in one of two identical waveguides, as illustrated in \figpanel{fig:Bloch}{a}. Such photons can be stored in optical fiber with transmission loss rates of less than 0.2 dB/km---or \SI{1}{\percent} per microsecond---at telecom wavelengths. On the microwave-frequency side, a dual-rail qubit is formed by a single excitation trapped in one of two stationary modes of microwave cavities~\cite{teoh_dual-rail_2023}, electromechanical resonators~\cite{wollack_quantum_2022,arrangoiz-arriola_resolving_2019,bozkurt_quantum_2023}, or qubits such as the transmon~\cite{koch_charge_2007, kubica_erasure_2023, Levine2024}. Regardless of the nature of the microwave dual-rail component, deterministic microwave gates exploiting Josephson junctions generate the microwave entanglement to be transferred to optics. Both cavities and qubits have attractive properties for building the microwave dual-rail component. First, superconducting cavities currently have intrinsic relaxation times of $\sim\SI{1}{\milli\second}$ and dephasing times of $\sim\SI{10}{\milli\second}$~\cite{teoh_dual-rail_2023}. Second, electromechanical resonators are developing as an alternative to microwave cavities. They may eventually provide advantages in lifetime, crosstalk, and compactness \cite{wollack_quantum_2022, arrangoiz-arriola_resolving_2019, bozkurt_quantum_2023,maccabe_nano-acoustic_2020}, yet are currently less mature. Third, transmons typically decohere faster than microwave cavities, with common relaxation and dephasing times of $\lesssim\SI{100}{\micro\second}$. They are a widely adopted and mature technology and have proven to be manufacturable on the scale needed for a single RSG~\cite{arute_quantum_2019, gong_quantum_2021,kim_evidence_2023}. The dephasing time for a dual-rail qubit based on flux-tunable transmons can be made comparable to that of cavities by a resonant coupling between the transmons~\cite{campbell_universal_2020, kubica_erasure_2023, Levine2024}. The resulting splitting of the two hybridized transmon modes is first-order insensitive to flux noise in the individual transmons, thus removing the largest source of dephasing for flux-tunable transmons while retaining their extraordinary versatility and fast two-qubit gates. For the sake of concreteness, we choose the transmon-based implementation of the microwave dual-rail component in the following analysis. 

A microwave dual-rail qubit is sketched in \figpanel{fig:Bloch}{b}, with two transmons serving as the rails for the dual-rail qubit and a coupler to control the interaction between the rails. The resulting logical subspace and leakage states are illustrated in \figpanel{fig:Bloch}{c}. The computational states of the microwave dual-rail qubit are the symmetric and antisymmetric single-excitation states. A Bloch-sphere representation of the dual-rail qubits is shown in \figpanel{fig:Bloch}{d}. We form a hybrid qubit by combining the microwave and optical dual-rail qubits. We motivate the introduction of the hybrid-qubit picture in the context of resource-state generation in Section~\ref{sec:protocol}.

\begin{figure}
    \centering
    \includegraphics[width = \linewidth]{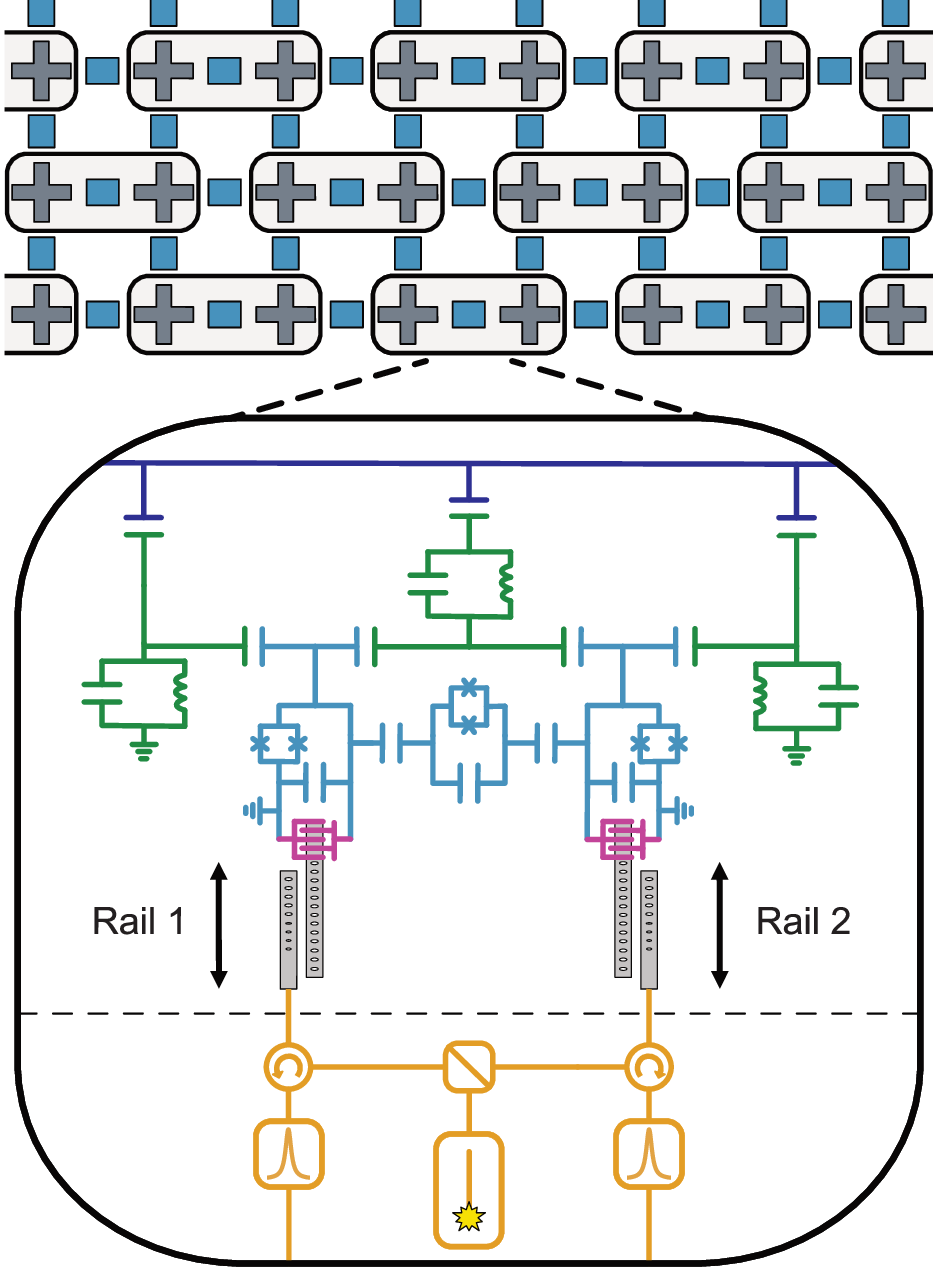}
    \caption{\textbf{Overview of the proposed architecture for generating resource states.} A superconducting quantum processor~\cite{arute_quantum_2019, gong_quantum_2021} is divided into dual-rail blocks. Each block contains two superconducting qubits and a tunable coupler (light blue). Tunable couplers also connect qubits belonging to different blocks. To each superconducting qubit we attach electrodes (purple) leading to the piezoelectric region of the microwave-optics transducer. Microwave resonators (green) are used for parity checks and read-out of the superconducting qubits through a read-out line (deep blue). The dashed line separates the on-chip components from the off-chip optical circuitry (orange) which consists of a pump laser (yellow star), a 50:50 beam splitter, circulators, and pump filters. The pump laser is split on the beam splitter, goes through a circulator into a single-side mirrored optical waveguide evanescently coupled to the piezo-optomechanical transducer (both grey). Optical photons leaving the transducer are directed away from the chip through an optical grating coupler (not shown) and a circulator. The pump is filtered out using e.g. a Fabry-P\'{e}rot optical filter, producing an optical dual-rail qubit.}
    \label{fig:network}
\end{figure}

The dual-rail qubit can be naturally integrated in large-scale superconducting quantum processors that use a square-grid layout~\cite{arute_quantum_2019,kosen_signal_2024}. In \figref{fig:network}, we show how we define dual-rail qubits on a square grid of transmon qubits in such a way that each dual-rail qubit in the bulk has six nearest neighbors. To turn the superconducting processor into an RSG, we attach a microwave-to-optics transducer to each transmon in each dual-rail qubit. We refer to a single dual-rail qubit with transducers as a block. The inset in \figref{fig:network} shows the circuit diagram of a single block. Each block initializes a dual-rail qubit by combining probabilistic microwave-optical spontaneous down-conversion with deterministic microwave gates. The optical photons are immediately routed to an external optical network using optical fiber. Although our scheme is compatible with any device capable of entangling optical and microwave photons~\cite{han_microwave-optical_2021}, we base the analysis on piezo-optomechanical transducers~\cite{meesala_non-classical_2024, mirhosseini_superconducting_2020, jiang_efficient_2020, jiang_optically_2023, chiappina_design_2023} for the sake of concreteness. These transducers support high entangling rates while dissipating little energy~\cite{jiang_efficient_2020,meesala_non-classical_2024}.

The piezo-optomechanical transducer couples an optical resonator to the transmon via mechanical motion. The frequency dependence of the optical resonator on mechanical motion introduces a three-wave-mixing nonlinearity that couples near-infrared ($\approx\SI{193}{\tera\hertz}$) optical modes to high-frequency ($\approx 5~\unit{\giga\hertz}$) mechanical modes. Without loss of generality, we focus on periodically patterned structures in silicon called optomechanical crystals (OMCs)~\cite{weaver_integrated_2023,jiang_optically_2023,meesala_non-classical_2024} to realize the optomechanical component. A strongly piezoelectric material such as lithium niobate is then used to coherently swap excitations between the mechanical mode and the transmon. An optical pump laser incident on the optical cavity bridges the energy gap between microwaves and optics and enhances the optomechanical interaction.

Placing the optical pump frequency $\omega_{\mathrm{L}}$ blue-detuned by the mechanical frequency $\omega_{\mathrm{m}}$ from the cavity frequency $\omega_\mathrm{o}$, we can realize a Hamiltonian that in the rotating frame of the laser takes the form $\hat{H} = \hat{H}_0 + \hat{H}_{\mathrm{int}}$~\cite{mirhosseini_superconducting_2020, zhong_microwave_2022}, with
\begin{eqnarray}
    \hat{H}_0 &=& \frac{1}{2}\hbar\omega_{\mathrm{q}}(t) \hat{\sigma}^z + \hbar\omega_{\mathrm{m}} \hat{b}^\dagger \hat{b} - \hbar\Delta\hat{c}^\dagger \hat{c}, \\
    \hat{H}_{\mathrm{int}} &=& \hbar g_{\mathrm{qm}}\mleft(\hat{\sigma}^+\hat{b} + \hat{\sigma}^-\hat{b}^\dagger\mright) + \hbar G(t) \mleft(\hat{b}^\dagger\hat{c}^\dagger + \hat{b}\hat{c}\mright).
    \label{eq:interaction_Hamiltonian}
\end{eqnarray}
Here, $\Delta=\omega_{\mathrm{L}} - \omega_{\mathrm{o}} = \omega_{\mathrm{m}}$ is the laser detuning, $\hat{\sigma}^z = \vert e\rangle\langle e\vert - \vert g\rangle\langle g\vert$ is the transmon's Pauli-\textit{Z} matrix,  $\hat{\sigma}^+$ ($\hat{\sigma}^-$) is the transmon raising (lowering) operator, $\hat{b}^\dagger$ ($\hat{b}$) is the creation (annihilation) operator for the mechanical mode, and $\hat{c}^\dagger$ ($\hat{c}$) is the creation (annihilation) operator for the optical mode. The transmon has a flux-tunable frequency $\omega_{\mathrm{q}}(t)$ and couples to the mechanical mode with coupling rate $g_{\mathrm{qm}}$. Thus, the transmon can be brought in and out of resonance with the mechanical mode to control the resonant swapping interactions. The optomechanical coupling $G(t) = g_{\mathrm{om}}\sqrt{n_{\mathrm{c}}(t)}$ is given by the single-photon optomechanical coupling rate $g_{\mathrm{om}}$ and the intracavity pump photon number $n_{\mathrm{c}}$. 

Previous proposals for heralding microwave-optical Bell pairs~\cite{zhong_proposal_2020, zhong_entanglement_2020} use a static, resonant coupling between a transducer and a superconducting resonator strongly coupled to a microwave waveguide. However, this configuration does not produce spectrally pure photons when the electrical and mechanical modes are strongly coupled, nor does it implement the multiphoton noise filtering that we propose in Section~\ref{sec:performance}. Therefore, the proposed architecture in \figref{fig:network}, which allows for a controllable electromechanical swap, is essential for the performance of an RSG.

\section{Protocol}
\label{sec:protocol}
\begin{figure}
    \centering
    \includegraphics[width=\linewidth]{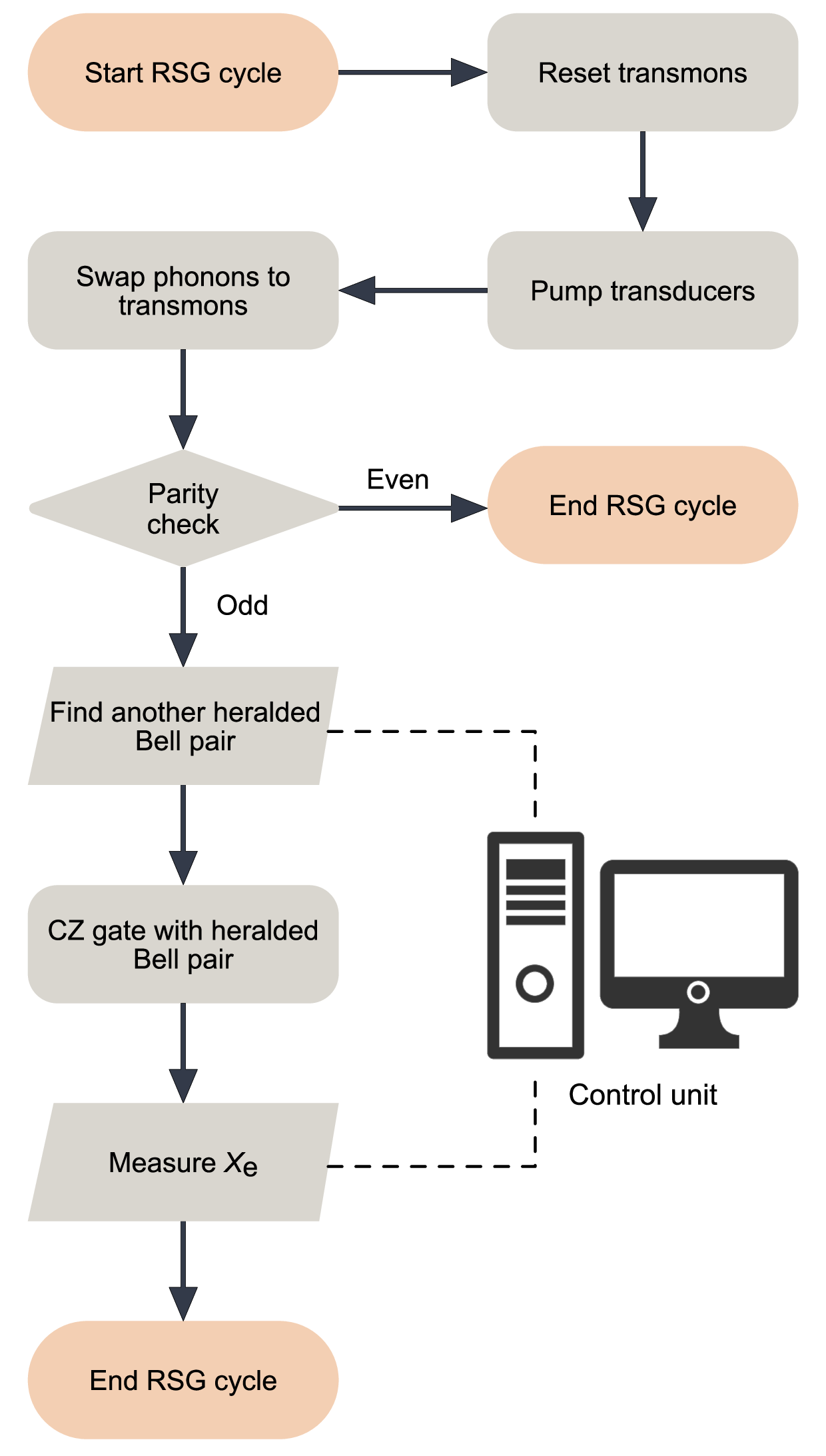}
    \caption{\textbf{The process flow in a single RSG cycle.}  Each block (\figref{fig:network}) undergoes the same set of instructions to herald microwave-optical Bell pairs and entangle them in a graph state. Entangled optical qubits are extracted from the graph state by the measurement of $X_e$ in each block. Oval shapes represent a start or end of a cycle, rectangular shapes represent a step to be executed, diamond shapes represent decisions, and parallelograms represent processes where classical information is either an input or output.}
    \label{fig:RSG_flow}
\end{figure}

We now explain how our architecture heralds microwave-optical Bell pairs and entangles them into graph states. Our protocol runs on a clock cycle, which we refer to as an RSG cycle. Each block in our architecture executes the same set of instructions every RSG cycle. \figref{fig:RSG_flow} shows the order of the steps that have to be executed. 
The RSG cycle time sets the rate at which resource states can be generated from a single superconducting processor. By building several copies of a single RSG, the total resource-state generation rate can be increased without increasing the complexity of control of a single RSG. With fusion-based quantum computing, the size of the resource state remains the same for an arbitrarily large computation---scaling simply becomes a question of how fast we can produce the resource states and how many RSGs we can afford to build. In principle, the resource state can consist of just four optical qubits~\cite{bartolucci_fusion-based_2023}, but bigger resource states tolerate more errors and are therefore desirable~\cite{bombin_fault-tolerant_2023}. 

The first step of an RSG cycle is an active reset~\cite{zhou_rapid_2021} on all transmons to force them to the ground state. The transmons are detuned away from the mechanical mode of the transducer to which they are coupled so that the electrical and mechanical modes are not yet hybridized, thus ensuring that the photons produced in the optomechanical down-conversion are spectrally pure. A single laser pulse then pumps the two transducers in each dual-rail qubit simultaneously and with equal intensity. In the idealized system, the optomechanical interaction produces the two-mode squeezed vacuum state 
\begin{equation}
    \vert\psi\rangle = \sqrt{1-\vert\lambda\vert^2}\sum\limits_{n=0}^\infty \lambda^n\vert n_{\mathrm{m}}\rangle\vert n_{\mathrm{o}}\rangle
    \label{eq:squeezed_state}
\end{equation}
in each transducer, where $\lambda$ is a squeezing parameter that quantifies the amount of optomechanical spontaneous down-conversion~\cite{hofer_quantum_2011} and $\vert n_{\mathrm{m(o)}}\rangle$ the $n$th Fock state of the mechanical (optical) mode. Because the optical cavity is strongly coupled to the external waveguide, the optical mode entangled with the mechanics is a traveling wave packet. For an open transducer system coupling to the environment, the state is represented by a density matrix $\hat{\rho}$ and the squeezing is given by the time integral of the optomechanical scattering rate $\Gamma_{\mathrm{om}} = 4\vert G(t)\vert^2/\kappa$, with $\kappa$ the optical linewidth~\cite{hofer_quantum_2011}. We return to the treatment of the open system in Section \ref{sec:performance}.  

The optical pump pulse prepares each transducer in the state $\vert \psi\rangle$. Next, the transmons are brought into resonance with the mechanical modes of their respective piezo-optomechanical transducer to swap the phonon to a photon in the transmon. The transmon frequency is tuned by more than $g_{\mathrm{qm}}$ in a time shorter than $1/g_{\mathrm{qm}}$ so that the swap has high fidelity. This requirement is satisfied with current transmon qubits given $g_{\mathrm{qm}}/(2\pi) \lesssim$ \SI{10}{\mega\hertz}~\cite{rol_fast_2019,arrangoiz-arriola_resolving_2019,weaver_integrated_2023,chiappina_design_2023}. The tunable coupler is set so that the transmons in the rails of the dual-rail qubit do not interact during the swap operation~\cite{Chen2014, Sung2021}. Once the swap is complete, the coupler can be used to turn the interaction between the transmons back on so that the logical qubit subspace is protected from dephasing in the individual transmons \cite{campbell_universal_2020,Levine2024}. 

The protocol proceeds with a parity check on the dual-rail microwave qubit. Several parity-check protocols apply to our scheme. For example, we can read out the state of a resonator that is symmetrically coupled to the two rail qubits~\cite{tornberg_high-fidelity_2010, frisk_kockum_undoing_2012, riste_deterministic_2013}, as shown in \figref{fig:network}. An odd outcome of the parity check heralds the excitation of one of the two rail transmons, but does not reveal which transmon has been excited, only that the dual-rail qubit is now inside the logical subspace [\figpanel{fig:Bloch}{b}]. Each microwave photon is accompanied by an optical photon having left the same transducer before the phonon-photon swap. Therefore, the result of the parity check is that the combined state of the two transmons and the two optical waveguides is projected onto the state 
\begin{equation}
    \vert \Psi^+\rangle = \mleft(\vert g0\rangle_{\mathrm{R1}}\vert e1\rangle_{\mathrm{R2}} + \vert e1\rangle_{\mathrm{R1}}\vert g0\rangle_{\mathrm{R2}}\mright)/\sqrt{2}.
\end{equation}
where we take the two optical drives to be in phase at their respective OMCs. Here, $\vert g0\rangle$ ($\vert e1\rangle$) is the state with the transmon in the ground (excited) state and no (one) optical photon in the attached optical waveguide. Crucially, events where no excitation is produced by the pump in either transducer lead to an even outcome in the parity check, as will events where both transmons are excited. 

\begin{figure}
    \centering
    \includegraphics[width = \linewidth]{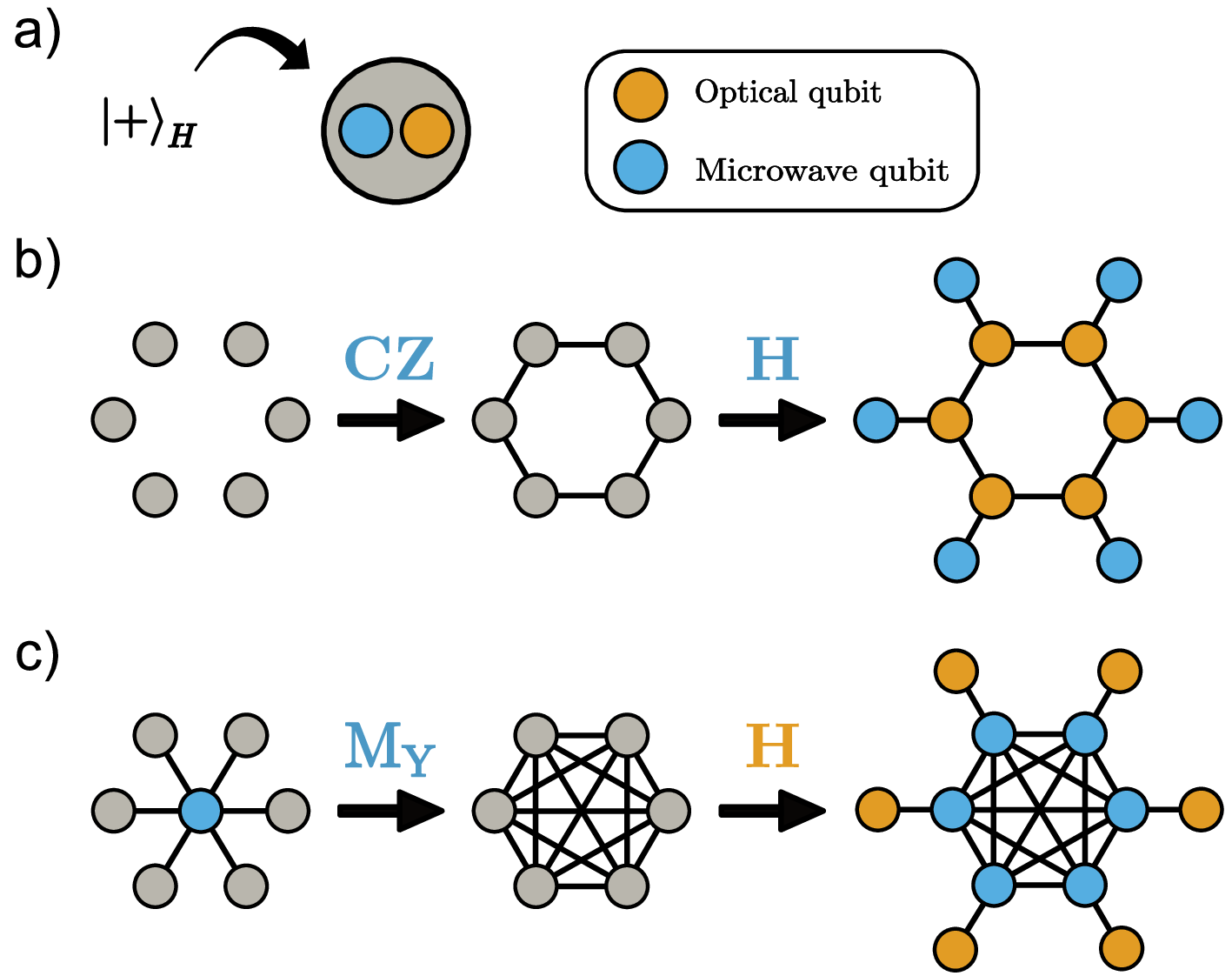}
    \caption{\textbf{Generation of microwave-optical resource states}. (a) The output from a single block is two entangled dual-rail qubits, which define a single hybrid qubit in the state $\vert + \rangle_L$. (b) To make a 6-ring resource state, we prepare six hybrid qubits (grey) and apply CZ gates between the microwave dual-rail qubits from each logical qubit. This produces a 6-ring resource state. A Hadamard gate (H) applied to all microwave qubits in the resource state pushes them out of their logical vertices. (c) A repeater graph useful for quantum communication can also be efficiently constructed using microwave CZ gates and measurements. Here, a measurement of the central microwave qubit in the $Y$ basis performs a local complementation of the graph, and a Hadamard gate applied to all optical qubits pushes the optical qubits out of the hybrid vertices.}
    \label{fig:Graphs}
\end{figure}

The heralded state $\vert \Psi^+\rangle$ is a Bell state of an optical dual-rail qubit and a microwave dual-rail qubit. Using the encoding (\figref{fig:Bloch})
\begin{subequations}
    \begin{align}
        \vert 0_\mathrm{e}\rangle &= \mleft(\vert g\rangle_{\mathrm{R1}}\vert e\rangle_{\mathrm{R2}} - \vert e\rangle_{\mathrm{R1}}\vert g\rangle_{\mathrm{R2}}\mright)/\sqrt{2}, \\
        \vert 1_\mathrm{e}\rangle &= \mleft(\vert g\rangle_{\mathrm{R1}}\vert e\rangle_{\mathrm{R2}} + \vert e\rangle_{\mathrm{R1}}\vert g\rangle_{\mathrm{R2}}\mright)/\sqrt{2}, \\
        \vert 0_\mathrm{o}\rangle &= \mleft(\vert 0\rangle_{\mathrm{R1}}\vert 1\rangle_{\mathrm{R2}} - \vert 1\rangle_{\mathrm{R1}}\vert 0\rangle_{\mathrm{R2}}\mright)/\sqrt{2}, \\
        \vert 1_\mathrm{o}\rangle &= \mleft(\vert 0\rangle_{\mathrm{R1}}\vert 1\rangle_{\mathrm{R2}} + \vert 1\rangle_{\mathrm{R1}}\vert 0\rangle_{\mathrm{R2}}\mright)/\sqrt{2},
    \end{align}
\end{subequations}
one can verify that
\begin{equation}
    \vert \Psi^+\rangle = \mleft(\vert 0_{\mathrm{e}}\rangle\vert 0_{\mathrm{o}}\rangle + \vert 1_{\mathrm{e}}\rangle\vert 1_{\mathrm{o}}\rangle\mright) / \sqrt{2}.
\end{equation}
The state $\vert \Psi^+\rangle$ can be viewed as a single qubit that has been redundantly encoded in both a microwave and optical qubit. Specifically, we define $\vert \Psi^+\rangle \equiv \vert +\rangle_H$ and treat the microwave-optical Bell pair as a single qubit prepared in the +1 eigenstate of the operator $\hat{X}_H = \hat{X}_\mathrm{e}\hat{X}_\mathrm{o}$. The operator $\hat{Z}_H$ has two equivalent representations: $\hat{Z}_H = \hat{Z}_\mathrm{e}$ and $\hat{Z}_H = \hat{Z}_\mathrm{o}$. The state $\vert +\rangle_H$ precesses at twice the coupling rate between the two transmons in the microwave dual-rail qubit. A graphical representation of the hybrid qubit redundantly encoded with one microwave qubit and one optical qubit is shown in \figpanel{fig:Graphs}{a}. 

To produce resource states, we prepare multiple blocks (\figref{fig:network}) in the state $\vert +\rangle_H$. By applying a CZ gate between microwave qubits from different blocks we can construct a graph state $G(E, V)$ of encoded hybrid qubits, defined as
\begin{equation}
    \vert G\rangle = \prod_{(i,j) \in E} \mathrm{CZ}_{i,j} \vert +\rangle_H^{\otimes V}.
\end{equation}
Equivalently, the graph state $\vert G\rangle$ of $k$ qubits can be described by its stabilizer generators,
\begin{equation}
    \hat{X}_{i, H} \prod\limits_{j \in \mathcal{N}(i)} \hat{Z}_{j,H},\quad \forall\ i \in \{1, 2, ..., k\},
\end{equation}
where $\mathcal{N}(i)$ is the set of vertices in the neighborhood of vertex $i$. The action of $\mathrm{CZ}_{i,j}$ on two hybrid qubits $i, j$ is 
\begin{equation}
    \hat{X}_{i, H} \rightarrow \hat{X}_{i,H}\hat{Z}_{j,H}, \quad \hat{X}_{j, H} \rightarrow \hat{Z}_{i,H}\hat{X}_{j,H}. 
\end{equation}
Using $\hat{X}_H = \hat{X}_{\mathrm{e}}\hat{X}_{\mathrm{o}}$ and $\hat{Z}_H = \hat{Z}_{\mathrm{o}}$, we see that application of CZ gates followed by measurement of the microwave dual-rail qubit in the $X$ basis implements $\hat{X}_{i, H} \rightarrow m_{i, \mathrm{e}} \hat{X}_{i,\mathrm{o}}\hat{Z}_{j,\mathrm{o}}$ and $\hat{X}_{j, H} \rightarrow m_{j, \mathrm{e}} \hat{Z}_{i,\mathrm{o}}\hat{X}_{j,\mathrm{o}}$, where $m = \pm 1$ is the measurement outcome. Thus, up to a Pauli frame correction given by the measurement outcomes, the quantum correlations that are left on the optical qubits are precisely those of a graph state. The microwave-optical hybrid qubits effectively allow us to teleport an arbitrary graph state prepared on the superconducting processor to optics using only single-qubit measurements.

As an example, \figpanel{fig:Graphs}{b} shows how we can construct a six-qubit hexagonal hybrid graph state (6-ring) represented by vertices $V$ and edges $E$. An all-optical 6-ring resource state can be extracted directly from the hybrid resource state by measuring all microwave qubits in the $X$ basis. Alternatively, we may apply a Hadamard gate on an optical or microwave dual-rail qubit to push this qubit out of its original vertex so that it forms a new vertex connected by an edge to its original vertex~\cite{hilaire_near-deterministic_2023}. This is illustrated in the final graph state in \figpanel{fig:Graphs}{b}, where we have applied a Hadamard gate on the microwave qubits. Measurement of the microwave qubits in the computational basis would remove them from the graph state, leaving an all-optical 6-ring that can be used in a fusion-network implementation of a surface code with linear optics~\cite{bartolucci_fusion-based_2023}. Other graph states, such as repeater graphs, can also be assembled deterministically, as illustrated in \figpanel{fig:Graphs}{c}.

A consequence of the probabilistic heralding of the microwave-optical Bell pairs is that many blocks must be pumped simultaneously to obtain multiple pairs to combine into a resource state. Because it is random which blocks produce a microwave-optical Bell pair in any given RSG cycle, the microwave processor that performs entangling gates between such pairs benefits from tunable coupling between many blocks. Thus, our scheme synergizes with efforts to improve superconducting qubit connectivity~\cite{stassi_scalable_2020, hazra_ring-resonator-based_2021,spring_high_2022}. However, our scheme does not require this. For example, blocks in \figref{fig:network} that did not herald a state $\vert +\rangle_H$ can be put in the state $\vert +_\mathrm{e}\rangle$ deterministically~\cite{Levine2024} and then used in the construction of the microwave-optical cluster state. 

\section{Performance analysis}
\label{sec:performance}

The primary application of our scheme is the production of optical graph states using a superconducting processor and \textit{imperfect} transducers. If transducers that could deterministically convert a microwave photon to an optical photon with negligible loss and added noise were available, the presented scheme would be less efficient than direct transduction of a graph state prepared on the superconducting processor. It is therefore of interest to investigate what level of performance would be required for our scheme and which metrics matter most to reach this level. 

\subsection{Squeezing and multiphoton noise}

A fundamental feature of two-mode squeezing is the production of states with two or more photons from a single laser pulse. Such states are unwanted and we refer to them as \textit{multiphoton noise}. This noise sets an upper bound on how hard we can squeeze the transducer, regardless of its imperfections such as thermal noise and photon loss, since the contribution of multiphoton states to $\vert \psi\rangle$ in \eqref{eq:squeezed_state} grows as we squeeze harder. However, squeezing too weakly implies that the heralding probability $p$ for a microwave-optical Bell state becomes prohibitively low. The size of the superconducting processor required for reliable generation of resource states is inversely proportional to $p$, and the need for long-range connections between transmons becomes more pressing for low values of $p$. Without access to a long-lived quantum memory, $p \lesssim\SI{1}{\percent}$ is too low. 

\begin{figure}
    \centering
    \includegraphics[width = \linewidth]{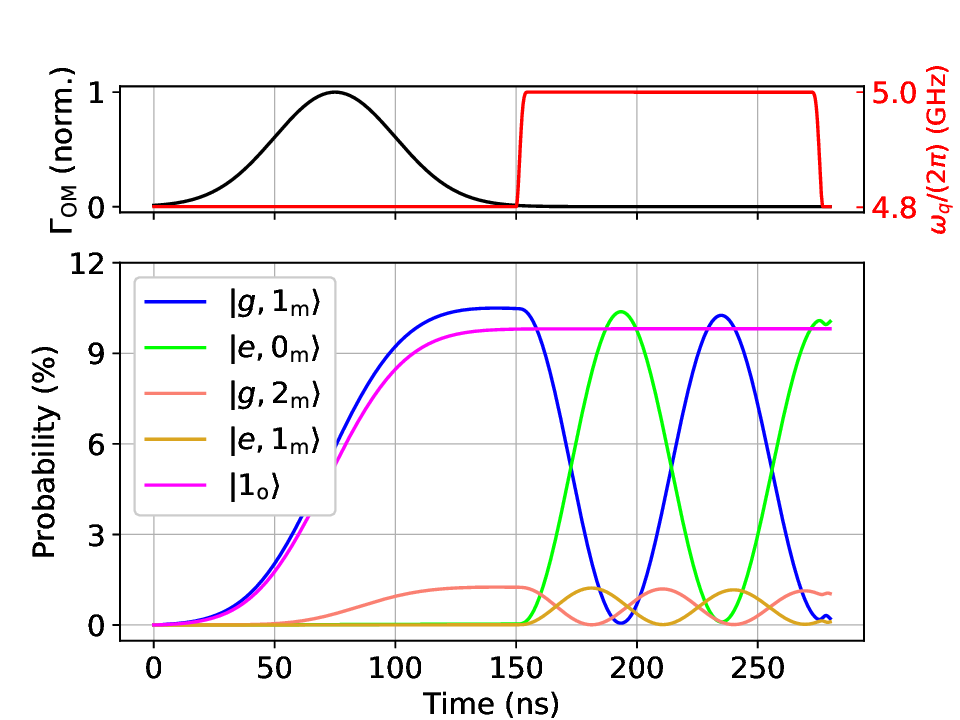}
    \caption{\textbf{Time-domain microwave-optical squeezing along with electromechanical swap and multiphoton noise filtering.}. Simulated populations (bottom) of the transmon and mechanical mode during the Gaussian pump pulse and subsequent swap to the transmon (top). Shown in pink is the probability of finding one optical photon in the corresponding optical wave packet leaving the transducer, after tracing out the transmon and mechanical mode. We give further details in Ref.~\cite{haug_supplementary_2023}.}
    \label{fig:triple_swap}
\end{figure}

In the protocol described in Section \ref{sec:protocol}, controlling the swap time between the transducer and the transmon allows us to take advantage of the different swap speeds of the one- and two-phonon states to suppress noise from the latter. A QuTiP simulation~\cite{johansson_qutip_2013} of a transmon and a piezo-optomechanical transducer during squeezing and subsequent swap operation is shown in \figref{fig:triple_swap}. Here, the two-phonon state is almost completely swapped back to the mechanical mode. At the same time, the one-phonon state is almost completely swapped to the transmon. This ensures that when the parity check heralds the state $\vert +\rangle_H$, the probability of finding two optical photons in the optical dual-rail qubit is suppressed. This allows us to squeeze the transducer harder without substantially increasing multiphoton noise in the heralded logical qubits until three-photon contributions become relevant. The microwave-optical heralding probability when pumping a transducer as shown in \figref{fig:triple_swap} is \SI{19}{\percent}.

\subsection{Hardware imperfections}

We now turn to the effects of non-ideal hardware. This will naturally depend on the choice of superconducting processor and transducer type used to implement the scheme. We continue to focus on transmon qubits and piezo-optomechanical transducers for concreteness. Our approach is to simulate a single rail in the dual-rail qubit using QuTiP~\cite{johansson_qutip_2013} as we pump the transducer and swap the phonons to the transmon. We construct the density matrix of a microwave-optical dual-rail qubit before the parity check by taking the tensor product of two of the simulated rails: $\hat{\rho}_{\mathrm{DR}} = \hat{\rho}_{\mathrm{R1}} \otimes \hat{\rho}_{\mathrm{R2}}$. We then perform the parity check by measuring $\hat{Z}_{\mathrm{q, R1}}\hat{Z}_{\mathrm{q, R2}}$, where $\hat{Z}_{\mathrm{q}}$ is the Pauli-$Z$ operator on the physical transmon qubit. Details about the simulation can be found in Ref.~\cite{haug_supplementary_2023}.

We analyze three parameter regimes for the piezo-optomechanical transducer detailed in Table \ref{tab:parameters}. First, we take the transducer design from Ref.~\cite{chiappina_design_2023} and find a Bell-state fidelity $F = \mathrm{Tr}\mleft(\hat{\rho}_{\mathrm{DR}}\vert +\rangle_H\langle +\vert\mright)$ of just \SI{38}{\percent} after heralding. Second, we run the same simulation with a transducer that has been better tailored to our scheme. In particular, we assume optical quality factors at half the value of state-of-the-art silicon photonic crystals~\cite{asano_photonic_2017}---roughly a quarter of the theoretical disorder-free scattering upper limit of a demonstrated OMC~\cite{ren_two-dimensional_2020}---as well as stronger coupling of the mechanical mode to the fridge bath and increased piezoelectric interaction strength. In this case, we find that the post-heralding fidelity increases to \SI{81}{\percent}. Third, we reduce the thermal noise in the transducer and find a fidelity of \SI{88}{\percent}. While they capture the overall quality of the generated optical states, these Bell-state fidelities do not describe the error mechanisms nor the error budget. In the following, we summarize the main sources of error and estimate the performance of each hardware component that would be necessary to reach the fault tolerance threshold. 
\begin{table*}
\caption{\label{tab:parameters}%
    \textbf{Parameters used in simulations of a piezo-optomechanical transducer during microwave-optical Bell-state preparation}. In addition to the design from Ref.~\cite{chiappina_design_2023}, we simulate using piezo-optomechanical transducers with improvements on three key parameters: optical quality factor, mechanical mode heating and electromechanical coupling. Optical quality factor and heating are the dominant factors determining the fidelity of the heralded Bell pairs. Electromechanical coupling has a much smaller impact on fidelity, but weak coupling leads to slower operation of the RSG and more Pauli errors. Parameters for the transmon qubit represent what can now be routinely manufactured in fixed-frequency transmon qubits~\cite{kosen_building_2022}.}
\begin{ruledtabular}
\begin{tabular}{l m{1.5cm} m{3.5cm} m{3.5cm} m{3.5cm}}
Parameter & Symbol & Current design~\cite{chiappina_design_2023} & Improved design & FBQC design   \\
\hline
Mechanical mode frequency & $\omega_{\mathrm{m}}$ & $2\pi \times \SI{5}{\giga\hertz}$ & $2\pi \times \SI{5}{\giga\hertz}$ & $2\pi \times \SI{5}{\giga\hertz}$\\
Electromechanical coupling & $g_{\mathrm{qm}}$ & $2\pi \times \SI{3}{\mega\hertz}$ & $2\pi \times \SI{6}{\mega\hertz}$ & $2\pi \times \SI{6}{\mega\hertz}$\\
Mechanical coupling to cold fridge bath & $\gamma_{0}$ & $2\pi \times \SI{10}{\kilo\hertz}$ & $2\pi \times \SI{100}{\kilo\hertz}$ & $2\pi \times \SI{100}{\kilo\hertz}$\\
Mechanical coupling to hot phonon bath & $\gamma_{\mathrm{b}}$ & $2\pi \times \SI{10}{\kilo\hertz}$ & $2\pi \times \SI{10}{\kilo\hertz}$ & $2\pi \times \SI{10}{\kilo\hertz}$\\ 
Mechanical dephasing rate & $\gamma_{\phi}$ & $2\pi \times \SI{10}{\kilo\hertz}$ & $2\pi \times \SI{10}{\kilo\hertz}$ & $2\pi \times \SI{10}{\kilo\hertz}$\\ 
Hot phonon bath turn-on rate* & $\gamma_{\mathrm{s}}$ & $2\pi \times \SI{215}{\kilo\hertz}$ & $2\pi \times \SI{215}{\kilo\hertz}$ & $2\pi \times \SI{215}{\kilo\hertz}$\\ 
Slow-growing fraction of hot phonon bath* & $\delta$ & 0.8 & 0.8 & 0.8\\ 
Steady-state phonon bath occupancy* & $n_{\mathrm{b}}$ & 20 & 5 & 1\\ 
Single-photon optomechanical coupling rate & $g_0$ & $2\pi \times \SI{830}{\kilo\hertz}$ & $2\pi \times \SI{830}{\kilo\hertz}$ & $2\pi \times \SI{830}{\kilo\hertz}$\\
Peak optomechanical coupling rate & $\max{G(t)}$ & $2\pi \times \SI{5.5}{\mega\hertz}$ & $2\pi \times \SI{9}{\mega\hertz}$ & $2\pi \times \SI{9}{\mega\hertz}$\\
Gaussian pump pulse duration (std. dev.) & $\tau_{\mathrm{pulse}}$ & \SI{25}{\nano\second} & \SI{25}{\nano\second} & \SI{25}{\nano\second}\\
Optical coupling to waveguide & $\kappa_{\mathrm{ex}}$ & $2\pi \times \SI{1}{\giga\hertz}$ & $2\pi \times \SI{1}{\giga\hertz}$& $2\pi \times \SI{1}{\giga\hertz}$\\
Intrinsic optical loss rate & $\kappa_{\mathrm{int}}$ & $2\pi \times \SI{400}{\mega\hertz}$ & $2\pi \times \SI{40}{\mega\hertz}$ & $2\pi \times \SI{40}{\mega\hertz}$\\
Transmon frequency (detuned) & $\omega_{\mathrm{q}}(0)$ & $2\pi \times  \SI{4.8}{\giga\hertz}$ & $2\pi \times  \SI{4.8}{\giga\hertz}$ & $2\pi \times  \SI{4.8}{\giga\hertz}$\\
Transmon frequency rise/fall time & $T_{r/f}$ & \SI{5}{\nano\second} & \SI{5}{\nano\second} & \SI{5}{\nano\second}\\
Transmon charging energy (in units of $h$) & $E_{\mathrm{C}}$ & \SI{200}{\mega\hertz} & \SI{200}{\mega\hertz} & \SI{200}{\mega\hertz}\\
Transmon energy decay time & $T_1$ & \SI{100}{\micro\second} & \SI{100}{\micro\second} & \SI{100}{\micro\second}\\
Transmon dephasing time & $T_{\phi}$ & \SI{100}{\micro\second} & \SI{100}{\micro\second} & \SI{100}{\micro\second}\\
Thermal photon bath occupancy & $n_a$ & 0.1 & 0.1 & 0.1\\ 
\hline
Simulated fidelity to $\vert +\rangle_H$ & $F$ & \textbf{38 \%} & \textbf{81 \%} & \textbf{88 \%}\\
\hline
\multicolumn{3}{l}{\small *Heating model for the optomechanical crystal in Ref.~\cite{meenehan_pulsed_2015}. See Ref.~\cite{haug_supplementary_2023} for details.} \\
\end{tabular}
\end{ruledtabular}
\end{table*}

Most dual-rail-qubit errors take the qubit out of the logical subspace. Such errors lead to qubit erasure and are relatively benign, with error-correction thresholds in fault-tolerant fusion networks and cluster states on the order of \SI{10}{\percent}~\cite{varnava_loss_2006, varnava_how_2008, bell_optimizing_2023}. In contrast, Pauli-error thresholds are currently only around \SI{1}{\percent}~\cite{bartolucci_fusion-based_2023, nickerson_measurement_2018}. Therefore, it is necessary to analyze the main sources of Pauli and erasure errors separately. 

\subsubsection{Pauli errors}

Pauli errors may arise from dephasing of the dual-rail qubit. Dual-rail qubits can have longer coherence times than their constituent superconducting qubits through hard-coded noise suppression~\cite{campbell_universal_2020,kubica_erasure_2023,Levine2024}. Decay and dephasing times of about \SI{1}{\milli\second} have been demonstrated for dual-rail qubits using transmon qubits with $T_1 \sim\SI{20}{\micro\second}$~\cite{Levine2024} and have not exhausted the limits of this technique. Another source of Pauli error is mechanical dephasing during the \SI{250}{\nano\second} before the swap to the superconducting qubit is complete (\figref{fig:triple_swap}). Cutting-edge OMCs support dephasing times of order \SI{100}{\micro\second}~\cite{maccabe_nano-acoustic_2020}. Close to such performance may be reached in piezo-optomechanical structures by, e.g., shrinking the piezoelectric section~\cite{chiappina_design_2023,jiang_optically_2023,meesala_non-classical_2024}. The average number of phonons in our heralding protocol is only $\sim0.1$, reducing the mechanical dephasing rate.

The parity check that heralds a microwave-optical Bell pair may give an erroneous result. This could lead to Pauli errors when we implement two-qubit gates on microwave dual-rail qubits that are outside the computational space. However, imperfect discrimination of the even- and odd-parity states will be detected when the dual-rail qubit is measured. We can then erase all affected qubits from the resource state before it is used in an implementation of a topological error-correcting code~\cite{bombin_logical_2023}. This is in contrast to proposals where the dual-rail qubits are used directly in circuit-based implementations of, e.g., the surface code~\cite{kubica_erasure_2023}. Moreover, fusion measurements on the resource states deterministically detect leakage, allowing us to treat leakage equivalently to loss. Pauli errors from measurement errors are strongly suppressed because both rails in the dual-rail qubit are measured. If the measurements are inconsistent, we similarly erase the qubit from the resource state.

\subsubsection{Erasure errors}

The two leading sources of erasure error are optical photon loss and mechanical thermal noise. Photons are lost due to absorption and scattering in the microwave-optics transducer. Intrinsic optical quality factors above 2 million can keep this error below \SI{10}{\percent} given common external coupling rates of 1 GHz. Piezo-optomechanical transducers~\cite{mirhosseini_superconducting_2020, meenehan_pulsed_2015, maccabe_nano-acoustic_2020, meesala_non-classical_2024} may reach this value given the order-of-magnitude improvements in optical quality that have already been demonstrated in similar photonic crystals~\cite{asano_photonic_2017}. So far, transducers have not explicitly been optimized with this goal in mind. 

Thermal noise in the mechanics or qubits can falsely herald the excitation of a Bell pair. This will produce the same state as a correctly heralded Bell pair followed by optical photon loss. Therefore, it is important that the mechanical mode is in its ground state at the beginning of each clock cycle, and that optical absorption ~\cite{maccabe_nano-acoustic_2020} does not heat up the mechanical mode appreciably before its phonons are swapped to the superconducting qubit. Pristine silicon OMCs heat up slowly enough that it can be tolerated on the timescale of Bell-state preparation~\cite{maccabe_nano-acoustic_2020, ren_two-dimensional_2020}. However, the addition of piezoelectric elements currently causes the mechanical mode to heat up within \SI{100}{\nano\second}~\cite{mirhosseini_superconducting_2020, chiappina_design_2023}. Designing the mechanical mode to be predominantly silicon-like may reduce the heating close to that of OMCs~\cite{chiappina_design_2023}. Qubit erasure errors from thermal noise can then be kept below \SI{10}{\percent}.
\begin{table*}
    \caption{\label{tab:overlaps}
     \textbf{Error budget for microwave-optical Bell state preparation.} Overlaps of the simulated state produced from a single block in our scheme with selected candidate states after heralding, and the observed error on the optical dual-rail qubits. Simulation parameters from Table~\ref{tab:parameters}.}
    \begin{ruledtabular}
    \begin{tabular}{lcccc}
    State & Current design~\cite{chiappina_design_2023} & Improved design & FBQC design & Error observed\\
    \hline
        $\vert +\rangle_L$ & 0.380 & 0.810 & 0.883 & N/A\\
        $\left(\vert g0\rangle_{\mathrm{R1}} \vert e0\rangle_{\mathrm{R2}} \pm \vert e0\rangle_{\mathrm{R1}}\vert g0\rangle_{\mathrm{R2}}\rangle\right)/\sqrt{2}$ & 0.614 & 0.130 & 0.059 & Photon loss\\
        $\left(\vert g0\rangle_{\mathrm{R1}} \vert e2\rangle_{\mathrm{R2}} \pm \vert e2\rangle_{\mathrm{R1}}\vert g0\rangle_{\mathrm{R2}}\rangle\right)/\sqrt{2}$ & 0.002 & 0.023 & 0.022& Multiphoton noise\\
        $\left(\vert g1\rangle_{\mathrm{R1}} \vert e1\rangle_{\mathrm{R2}} \pm \vert e1\rangle_{\mathrm{R1}}\vert g1\rangle_{\mathrm{R2}}\rangle\right)/\sqrt{2}$ & 0.001 & 0.010 & 0.010 & Multiphoton noise\\
        $\left(\vert g0\rangle_{\mathrm{R1}} \vert e3\rangle_{\mathrm{R2}} \pm \vert e3\rangle_{\mathrm{R1}}\vert g0\rangle_{\mathrm{R2}}\rangle\right)/\sqrt{2}$ & $<0.001$ & 0.010 & 0.011 & Multiphoton noise\\
        $\left(\vert g2\rangle_{\mathrm{R1}} \vert e1\rangle_{\mathrm{R2}} \pm \vert e1\rangle_{\mathrm{R1}}\vert g2\rangle_{\mathrm{R2}}\rangle\right)/\sqrt{2}$ & $<0.001$ & 0.008 & 0.009 & Multiphoton noise\\
        $\left(\vert g0\rangle_{\mathrm{R1}} \vert e1\rangle_{\mathrm{R2}} - \vert e1\rangle_{\mathrm{R1}}\vert g0\rangle_{\mathrm{R2}}\rangle\right)/\sqrt{2}$ & 0.001 & 0.006 & 0.005& Phase flip\\   
    \end{tabular}
    \end{ruledtabular}
\end{table*}

\subsection{Error hierarchy}

As a concrete example of how hardware imperfections cause errors on the microwave-optical dual-rail qubits, Table~\ref{tab:overlaps} shows the breakdown of the simulated density matrices $\rho_{\mathrm{DR}}$ after the parity check in terms of overlap with states that constitute errors on the qubits. The simulated error rates can be reduced by optimizing the squeezing strength to suit the performance of the transducer, particularly for the improved design, where multiphoton noise is significant despite the proposed noise filtering (\figref{fig:triple_swap}). The observed photon loss rate using the best-performing transducer in Table~\ref{tab:overlaps} is below the threshold set by fusion erasure for FBQC with 6-ring resource states~\cite{bartolucci_fusion-based_2023}, but the threshold must be increased further to tolerate multiphoton noise, Pauli errors, and fusion failure. This can be achieved with loss-tolerant encoding~\cite{bartolucci_fusion-based_2023, bell_optimizing_2023} and dynamic bias arrangement~\cite{bombin_increasing_2023}.

\begin{figure}
    \centering
    \includegraphics[width=\linewidth]{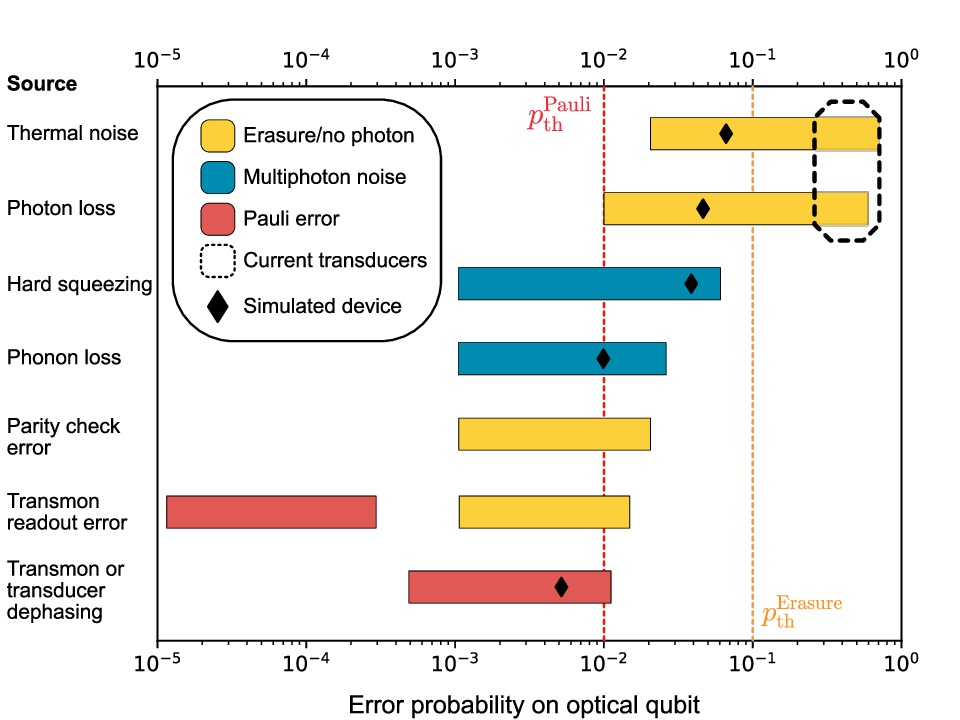}
    \caption{\textbf{Error hierarchy of the entangled optical qubits produced by the RSG.} The bars represent feasible error rates from the dominant error sources, from pessimistic estimates based on currently available transducers to optimistic estimates based on, e.g., theoretical disorder-free performance limits of current optomechanical crystals~\cite{ren_two-dimensional_2020} and experimentally demonstrated silicon photonic crystals~\cite{asano_photonic_2017}. We have indicated the performance of the simulated improved transducer ($F = \SI{81}{\percent}$) in Table~\ref{tab:parameters}. The simulation does not give the source of the observed errors, and therefore the indicated values are associated with some uncertainty. The parity check in our simulation is error-free and therefore not indicated. This has little effect on the overall error on the qubit because errors from microwave thermal noise and optical photon loss dominate. Transmon readout is not part of our simulation but is also expected to be a minor contribution to both erasure and Pauli errors. Dashed lines represent known fault-tolerance thresholds using error models for linear optical quantum computation~\cite{bartolucci_fusion-based_2023, nickerson_measurement_2018, bell_optimizing_2023, bombin_increasing_2023} and should be interpreted as upper bounds on the thresholds for FBQC using the proposed RSGs. Multiphoton noise will contribute predominantly to erasure, but a fraction could also be turned into Pauli error if photons are lost at a later stage.}
    \label{fig:error_hierarchy}
\end{figure}

Dual-rail microwave-optical qubits have an error hierarchy very similar to that found for all-microwave dual-rail qubits~\cite{teoh_dual-rail_2023, kubica_erasure_2023}. In \figref{fig:error_hierarchy}, we give an overview of feasible hardware performance in terms of errors on the optical qubits produced by the RSG, ranging from pessimistic with little improvement over current devices, to optimistic with a transducer that combines state-of-the-art performance for the transmon, optical quality factor similar to all-optical silicon photonic crystals ~\cite{asano_photonic_2017} and thermal noise levels of state-of-the-art optomechanical crystals. 

\subsection{Quasiparticle poisoning of superconductors}
We conclude the performance analysis with a note on the integration of superconducting circuits and optics. Stray optical pump photons can excite quasiparticles in superconductors that temporarily reduce the lifetime of transmon qubits~\cite{mirhosseini_superconducting_2020, benevides_quasiparticle_2023}. While our scheme does not rely on the preservation of a transmon's quantum state during the optical pulse, it is important that the transmons operate with high coherence within $\sim \SI{10}{\nano\second}$ after the pump pulse. Otherwise, swapping of the mechanical state to the superconducting qubit is not possible. Using only superconductors with short quasiparticle lifetimes, such as NbTiN~\cite{ meesala_non-classical_2024, jiang_optically_2023}, or shielding the superconducting qubits from optical radiation through, e.g., infrared filtering, may address this issue. Gap engineering of the Josephson junction in aluminum transmons was recently found to increase the tolerance to optical illumination by over three orders of magnitude~\cite{mcewen_resisting_2024}. We expect that combining gap engineering of the aluminum Josephson junction with low-quasiparticle-lifetime materials for the rest of the microwave circuitry will yield sufficiently light-insensitive qubits to run the protocols described here.  

\section{Fault-tolerance overhead}

Fault-tolerant quantum technologies typically come with a steep cost in overhead once the performance of the hardware can be kept below the relevant thresholds. In general, the overhead needed to reach a target performance level depends on how far below the threshold the hardware is performing. Finding precise error thresholds for FBQC with microwave-optical RSGs is beyond the scope of this proposal. In this section, we compare the overhead of FBQC with microwave-optical RSGs to other candidate fault-tolerant schemes, assuming that the performance is just barely below the fault-tolerance threshold for all schemes. The aim is to estimate, based on the fundamental working principles of each approach, how costly FBQC with our RSGs is compared to established alternatives.

\subsection{Comparison to all-optical schemes}

To illustrate how efficient our scheme can be for producing optical resource states, we compare the number of blocks needed to produce the 6-ring [\figpanel{fig:Graphs}{b}] for a fault-tolerant fusion network~\cite{bartolucci_fusion-based_2023} to the equivalent number of optical photons required using linear optics. While this is an open research problem in classical and quantum photonics, we assume that for the latter case a fast and low-loss optical switch can be produced so that optical photons can be multiplexed---which is required for an all-optical approach~\cite{rudolph_why_2017, bonneau_effect_2015}. A 6-ring can be made from three copies of a 3-qubit linear cluster state, which is locally equivalent to a Greenberger-Horne-Zeilinger (GHZ) state, using type-I fusion~\cite{browne_resource-efficient_2005}. All-optical circuits can ballistically produce GHZ states from six photons with probability $1/2^5$ with perfect photon number-resolving detectors~\cite{varnava_how_2008}. Therefore, the average number of photons needed to produce a GHZ state is 192. Fusing three such states using type-I fusion gates, each with probability of success 1/2, brings the average number of photons needed to make a 6-ring up to 4608 if states that underwent failed fusion are discarded. In contrast, our scheme would not need extra photons to entangle qubits and the entangling gates are deterministic. The number of microwave-optical Bell pairs required to create a 6-ring is six, but because each block produces a Bell pair with limited probability $p$, the number of blocks needed to reliably produce such states is $2\times6/p$, where $p$ is the probability of heralding a microwave-optical Bell pair from a single block. We estimate that $p$ can reach \SI{20}{\percent} using the scheme and architecture presented in this work. Higher heralding probabilities would likely generate too much multiphoton noise from the strong squeezing required. 

The 6-ring or other resource states for fusion networks will likely need to be concatenated with a graph code to increase tolerance to photon loss and fusion failure~\cite{bell_optimizing_2023}. As the size of the resource state is scaled up, the overhead implied by the probabilistic linear optical generation of such states increases rapidly. With our scheme, the overhead scales linearly with the size of the resource state. Moreover, our scheme circumvents the need for large-scale switching of photons for resource-state generation. With lossy switches, minimizing the switch depth is crucial. 

Besides the favorable scaling with resource state size compared to linear optics, the utility of our scheme also depends on its repetition rate. Current piezo-optomechanical transducers are limited by heating to order \SI{10}{\kilo\hertz} rates~\cite{meesala_non-classical_2024,chiappina_design_2023,jiang_optically_2023}. Optomechanical crystals with improved thermal anchoring are under investigation to address this issue~\cite{ren_two-dimensional_2020, kolvik_clamped_2023}. Several other transducer types, each with their benefits and challenges, are under development~\cite{jiang_efficient_2020,mirhosseini_superconducting_2020,delaney_superconducting-qubit_2022,jiang_optically_2023,weaver_integrated_2023,sahu_entangling_2023,meesala_non-classical_2024,han_microwave-optical_2021}. Improved transducers may relax or even completely remove the heating constraint, allowing resource-state generation rates to approach the limit set by the microwave gate times---estimated at order \SI{1}{\mega\hertz}. Such repetition rates would make a single RSG cycle similar to a surface code cycle on a traditional superconducting processor. 

\subsection{Comparison to all-microwave schemes}
For long-term applications, we now estimate how many RSGs would be required to match the computational power of a superconducting processor on its own. Using superconducting qubits to build RSGs rather than a surface-code processor will incur an overhead because the heralding of the microwave-optical Bell pairs is probabilistic and noisy. This overhead could leave the modular approach presented here at a disadvantage compared to scaling the size of the superconducting processor to hundreds of thousands of qubits, assuming this could be done without optics. We argue that constructing a fusion-based quantum computer is not as costly as it might appear thanks to the power of low-loss optical fiber delay lines. In a surface code of distance $d$, the number of logical qubits that can be encoded with one RSG is~\cite{litinski_how_2023} 
\begin{equation}
    N_L = \frac{\Gamma_{\mathrm{RSG}} t_{\mathrm{d}}}{d^2},
\end{equation}
where $\Gamma_{\mathrm{RSG}}$ is the resource-state generation rate and $t_\mathrm{d}$ is the maxmimum delay time from when a resource state is generated until all its qubits have been measured. If $t_\mathrm{d} = \SI{10}{\micro\second}$, then optical qubits stored in commercially available fiber with a loss rate of 0.16 dB/km~\cite{morana_extreme_2020} only suffer \SI{7}{\percent} additional loss. This extra delay loss only affects a few photons, and the delay loss can therefore exceed the threshold for baseline loss---i.e., the photon loss considered in this paper---without lowering this threshold substantially~\cite{bombin_interleaving_2021}. We assume for the purpose of illustration that it takes 100 dual-rail qubits to build one RSG and that each RSG cycle takes \SI{1}{\micro\second}. Then $N_L = 10/d^2$, and each superconducting dual-rail qubit contributes $1/10d^2$ towards a logical qubit. The equivalent number for a dual-rail qubit on a superconducting processor using the rotated surface code is $1/2d^2$, so the additional overhead for our proposal in this example is a factor five. This is before accounting for the overhead reductions that LDPC codes and an active-volume architecture might bring. 

A future full-stack comparison could also take into account that microwave systems do not easily scale beyond about a 1000 qubits~\cite{krinner_engineering_2019} in a single cryogenic system with current schemes. Thus, optics is likely needed either way to scale up microwave qubits to fault tolerance through optically heralded microwave entanglement \cite{krastanov_optically_2021,duan_long-distance_2001}. Alternative microwave approaches using tailored delay lines \cite{ferreira_deterministic_2024} and LDPC codes \cite{bravyi_high-threshold_2024} are, similar to microwave-optics devices, in an early stage of development. Their advantage lies in deterministic entanglement generation \cite{ferreira_deterministic_2024}, and their challenge in long-range connectivity \cite{bravyi_high-threshold_2024}. Optics has long been the preferred information carrier in classical non-local interconnects and is taking over from microwaves in datacenters as well as between and eventually within electronic chips. One may expect this trend to continue both within and between quantum processors of any kind.

\section{Conclusion}

We propose a scheme for generating microwave-optical cluster states of dual-rail encoded qubits. Our architecture uses imperfect quantum transducers as sources of microwave-optical Bell pairs. Next, it deterministically entangles Bell pairs into cluster states. Single-qubit measurements of the microwave qubits reduce the hybrid cluster states to all-optical resource states, thus teleporting an arbitrary graph state prepared on the superconducting processor to optics. Deterministic entanglement generation in the microwave domain allows our scheme to produce optical resource states faster than equivalent schemes in linear optics, even if individual optical photons could be generated at orders-of-magnitude higher rates. Although the analysis focuses on generating entangled optical photons from a microwave processor, this work is only a first step in exploring the landscape of architectures harnessing hybrid microwave-optical qubits. The proposed scheme can be used to turn microwave-frequency superconducting processors into resource-state generators useful for fault-tolerant quantum communication, computation, and sensing. Improved isolation of qubits from stray light would place the scheme within reach of proof-of-principle demonstrations with current hardware.

Research into hybrid microwave-optics devices is accelerating. If they continue to improve on their current steep path, and are tailored to the proposed scheme, we find that fault-tolerance thresholds are within reach and can eventually be surpassed. This would enable combining the exquisite quantum control of microwave circuits with the non-local connectivity of optics. The scheme is competitive with the low-connectivity all-microwave approach even when the constraint of a single cryogenic system is hypothetically removed. Initially, the benefits of this approach will likely be felt most in situations where the leading information carriers are optical photons either way. This includes long-distance networking, optical quantum sensing and computing, and scaling up microwave processors well beyond one cryogenic system.

\begin{acknowledgments}

\paragraph*{Acknowledgements.}
We thank Giulia Ferrini, Johan Kolvik, Paul Burger, Joey Frey, Simone Gasparinetti, Per Delsing, and Terry Rudolph for helpful discussions. We acknowledge support from the Knut and Alice Wallenberg foundation through the Wallenberg Centre for Quantum Technology (WACQT), from the European Research Council via Starting Grant 948265, and from the Swedish Foundation for Strategic Research.
\paragraph*{Contributions.} T.H.H. led the theoretical and numerical analysis. T.H.H. conceived of the approach in collaboration with R.V.L.; A.F.K. provided input at a later stage. A.F.K. and R.V.L. provided support on the analysis. All authors contributed to the writing with T.H.H. in the lead. R.V.L. conceived and supervised the project. 

\end{acknowledgments}

\bibliography{RSGbib}
\newpage
\renewcommand{\theequation}{S.\arabic{equation}}
\renewcommand{\thefigure}{S\arabic{figure}}
\renewcommand{\thetable}{S\Roman{table}}
\renewcommand{\thesection}{S\arabic{section}}
\setcounter{equation}{0}
\setcounter{figure}{0}
\setcounter{section}{0}

\onecolumngrid
\begin{center}
    \large{\textbf{Supplemental Material for: "Quantum Resource States from Heralded Microwave-Optical Bell Pairs"}}
\end{center}

\section{Two-mode squeezing of an open optomechanical system}

We investigate the optomechanical squeezing responsible for creating the entangled phonon-photon pair that we use to create a microwave-optical Bell pair. During the pump pulse, the mechanical mode is dispersively coupled to the qubit. The Hamiltonian of the system can then be approximated~\cite{aspelmeyer_cavity_2014,mirhosseini_superconducting_2020} as $\hat{H} = \hat{H}_0 + \hat{H}_{\mathrm{int}}$ with 
\begin{eqnarray}
    \hat{H}_0 &=& \frac{1}{2}\hbar\omega_\mathrm{q}(0) \hat{\sigma}^z + \hbar\omega_\mathrm{m} \hat{b}^\dagger \hat{b} + \hbar \chi \hat{b}^\dagger \hat{b}\hat{\sigma}^z - \hbar\omega_\mathrm{m} \hat{c}^\dagger\hat{c}, \\
    \hat{H}_{\mathrm{int}} &=& \hbar G(t) \mleft(\hat{b}^\dagger\hat{c}^\dagger + \hat{b}\hat{c}\mright).
\end{eqnarray}
Here, $\hbar$ is the reduced Planck constant, $\omega_\mathrm{q}(0)$ is the qubit frequency before the swap with the mechanical mode, $\omega_\mathrm{m}$ is the frequency of the mechanical mode, $G(t)$ is the optomechanical coupling strength, $\hat{\sigma}^z = \vert e\rangle\langle e\vert - \vert g\rangle\langle g\vert$ is the Pauli Z matrix,  $\hat{\sigma}^+$ ($\hat{\sigma}^-$) is the qubit raising (lowering) operator, $\hat{b}^\dagger$ ($\hat{b}$) is the mechanical mode creation (annihilation) operator, and $\hat{c}^\dagger$ ($\hat{c}$) is the the optical mode creation (annihilation) operator. We have defined the mechanical frequency shift depending on the state of the qubit,
\begin{equation}
    \chi = -\frac{g^2_{\mathrm{qm}} E_C/\hbar}{\Delta\left(\Delta-E_C/\hbar\right)} \ll \omega_m,
\end{equation}
with $E_C$ the charging energy of the superconducting qubit and $\Delta = \omega_\mathrm{q}(0)-\omega_\mathrm{m} \gg g_{\mathrm{qm}}$ the qubit-mechanics detuning~\cite{koch_charge_2007, blais_circuit_2021}.

We assume that the qubit remains in the ground state at all times during the pump pulse such that the qubit state and the optomechanical state factorize, $\rho(t) = \rho_q(0) \otimes \rho_{\mathrm{OM}}(t)$. Absorbing the qubit-induced mechanical frequency shift $\omega_m - \chi/2 \rightarrow \omega_m$ and moving to a frame rotating at the mechanical frequency, the Heisenberg-Langevin equations for the optomechanical system are
\begin{eqnarray}
    \Dot{\hat{c}} &=& -\frac{\kappa}{2}\hat{c} + iG(t) \hat{b}^\dagger + \sqrt{\kappa} \hat{c}_{\mathrm{in}},
    \label{eq:dot_c} \\
    \Dot{\hat{b}} &=& -\frac{\gamma}{2}\hat{b} + iG(t) \hat{c}^\dagger + \sqrt{\gamma}\hat{b}_{\mathrm{in}}.
    \label{eq:dot_b_1}
\end{eqnarray}
Here, $\kappa$ is the total optical linewidth and $\gamma$ is the total mechanical linewidth. We have defined the standard input fields~\cite{gardiner_input_1985} to the optical and mechanical modes as $\hat{c}_{\mathrm{in}}$ and $\hat{b}_{\mathrm{in}}$, respectively. The input operators in Eqs.~(\ref{eq:dot_c}) and (\ref{eq:dot_b_1}) obey the standard commutation relations~\cite{gardiner_input_1985}
\begin{equation}
    \mleft[\hat{b}_{\mathrm{in}}(t), \hat{b}^\dagger_{\mathrm{in}}(t')\mright] = \mleft[\hat{c}_{\mathrm{in}}(t), \hat{c}^\dagger_{\mathrm{in}}(t') \mright] = \delta(t-t').
\end{equation}
We exploit the fact that $\kappa \gg G(t)$ at all times to approximate the optical response as instantaneous, setting $\Dot{\hat{c}} = 0$ in \eqref{eq:dot_c}. This yields the coupled equations
\begin{eqnarray}
    \hat{c} &=& i\sqrt{\frac{\Gamma_{\mathrm{OM}}}{\kappa}}\hat{b}^\dagger + \frac{2}{\sqrt{\kappa}}\hat{c}_{\mathrm{in}}, \\
    \Dot{\hat{b}} &=& -\frac{\gamma - \Gamma_{\mathrm{OM}}}{2}\hat{b} + i\sqrt{\Gamma_{\mathrm{OM}}} \hat{c}_{\mathrm{in}}^\dagger + \sqrt{\gamma}\hat{b}_{\mathrm{in}},
    \label{eq:dot_b_2}
\end{eqnarray}
where $\Gamma_{\mathrm{OM}}(t) = 4G(t)^2/\kappa$ is the optomechanical scattering rate. We can solve \eqref{eq:dot_b_2} analytically for the case of a square pump pulse of duration $\tau$. Direct integration over the pump pulse yields
\begin{equation}
    \hat{b}(\tau) = e^{\frac{\Gamma_{\mathrm{OM}}-\gamma}{2}\tau}\hat{b}(0)+i\sqrt{\Gamma_{\mathrm{OM}}}e^{\frac{\Gamma_{\mathrm{OM}}-\gamma}{2}\tau}\int\limits_0^\tau dt\ e^{-\frac{\Gamma_{\mathrm{OM}}-\gamma}{2}t} \hat{c}_{\mathrm{in}}(t) + \sqrt{\gamma}e^{\frac{\Gamma_{\mathrm{OM}}-\gamma}{2}\tau}\int\limits_0^\tau dt\ e^{-\frac{\Gamma_{\mathrm{OM}}-\gamma}{2}t} \hat{b}_{\mathrm{in}}(t).
\end{equation}
We now write $\gamma = \alpha \Gamma_{\mathrm{OM}}$. Optomechanical devices are often operated in the high-cooperativity regime, where $\mathcal{C}=\Gamma_{\mathrm{OM}}/\gamma \gg 1$ (or equivalently $\alpha \ll 1$)~\cite{aspelmeyer_cavity_2014,meenehan_pulsed_2015, maccabe_nano-acoustic_2020} such that the coherent read-out occurs faster than the mechanical decoherence. However, for our scheme we are more likely to be in the regime where $\alpha \sim 1$ because we want phonons to leak out before starting the next clock cycle of our protocol. By defining the operators~\cite{hofer_quantum_2011}
\begin{eqnarray}
    \hat{C}_{\mathrm{in}} &=& \sqrt{\frac{\left(1-\alpha\right)\Gamma_{\mathrm{OM}}}{1-e^{(\alpha-1)\Gamma_{\mathrm{OM}}\tau}}}\int\limits_0^\tau dt\ e^{-\frac{(1-\alpha)\Gamma_{\mathrm{OM}}t}{2}} \hat{c}_{\mathrm{in}}(t), \\
    \hat{C}_{\mathrm{out}} &=& \sqrt{\frac{\left(1-\alpha\right)\Gamma_{\mathrm{OM}}}{e^{\left(1-\alpha\right)\Gamma_{\mathrm{OM}}\tau}-1}}\int\limits_0^\tau dt\ e^{\frac{(1-\alpha)\Gamma_{\mathrm{OM}t}}{2}} \hat{c}_{\mathrm{out}}(t),
\end{eqnarray}
and using the well-known input-output relation $\hat{c}_{\mathrm{out}} = -\hat{c}_{\mathrm{in}} + \sqrt{\kappa}\hat{c} $, we have to lowest order in $\Gamma_{\mathrm{OM}}\tau$:
\begin{eqnarray}
    \hat{C}_{\mathrm{out}} &=& e^{\frac{(1-\alpha)\Gamma_{\mathrm{OM}}\tau}{2}} \hat{C}_{\mathrm{in}} + \frac{i}{\sqrt{1-\alpha}}\sqrt{e^{\left(1-\alpha\right)\Gamma_{\mathrm{OM}}\tau}-1} \hat{b}^\dagger(0),
    \label{eq:squeeze1} \\
    \hat{b}(\tau) &=& e^{\frac{(1-\alpha)\Gamma_{\mathrm{OM}}\tau}{2}}\hat{b}(0) + \frac{i}{\sqrt{1-\alpha}}\sqrt{e^{\left(1-\alpha\right)\Gamma_{\mathrm{OM}}\tau}-1} \hat{C}_{\mathrm{in}}^\dagger + \hat{F}.
    \label{eq:squeeze2}
\end{eqnarray}
The operator
\begin{equation}
    \hat{F} = \sqrt{\gamma}e^{\frac{(1-\alpha)\Gamma_{\mathrm{OM}}}{2}\tau}\int\limits_0^\tau dt\ e^{-\frac{(1-\alpha)\Gamma_{\mathrm{OM}}}{2}t} \hat{b}_{\mathrm{in}}(t)
\end{equation}
represents the noise coming from the coupling to the thermal phonon bath. To lowest order in $\Gamma_{\mathrm{OM}}\tau$, we have $\left[\hat{C}_{\mathrm{in}}, \hat{C}_{\mathrm{in}}^\dagger\right] = \left[\hat{C}_{\mathrm{out}},\hat{C}_{\mathrm{out}}^\dagger\right] = 1$ and $\left[\hat{F}, \hat{F}^\dagger\right] = \gamma\tau$. Our scheme \textit{requires} $\Gamma_{\mathrm{OM}}\tau \ll 1$ to keep the probability of exciting multiple phonon-photon pairs low, so keeping only lowest-order terms is a good approximation.

In the absence of mechanical noise, we see that in the limit $\gamma \rightarrow 0$ ($\alpha \rightarrow 0$), corresponding to high mechanical quality factors, Eqs.~(\ref{eq:squeeze1}) and (\ref{eq:squeeze2}) represent two-mode squeezing of the mechanical mode and the output optical field with $e^{\Gamma_{\mathrm{OM}}\tau/2} = \cosh(r)$ and $\sqrt{e^{\Gamma_{\mathrm{OM}}\tau}-1} = \sinh(r)$. However, even when $\alpha \approx 1$ we get a similar behavior. We have
\begin{equation}
    \lim\limits_{\alpha \rightarrow 1} \sqrt{\frac{e^{\left(1-\alpha\right)\Gamma_{\mathrm{OM}}\tau}-1}{1-\alpha}} = \sqrt{\Gamma_{\mathrm{OM}}\tau},
\end{equation}
and with $\Gamma_{\mathrm{OM}}\tau \lesssim 10^{-1}$, the relevant order of magnitude for our scheme, we obtain
\begin{eqnarray}
    \hat{C}_{\mathrm{out}} &=& \hat{C}_{\mathrm{in}} + i\sqrt{\Gamma_{\mathrm{OM}}\tau} \hat{b}^\dagger(0), \\
    \hat{b}(\tau) &=& \hat{b}(0) + i\sqrt{\Gamma_{\mathrm{OM}}\tau}\hat{C}_{\mathrm{in}}^\dagger,
\end{eqnarray}
to order $\sqrt{\Gamma_{\mathrm{OM}}\tau}$, regardless of whether $\alpha$ is close to unity or much smaller than unity. If $\alpha \gg 1$ (i.e., $\gamma \gg \Gamma_{\mathrm{OM}}$), then we no longer have something that resembles two-mode squeezing because the phonons in the mechanical mode escape faster than we can produce them.

In the above, we neglected thermal noise in the mechanics. Such noise would increase the mechanical dephasing rate and increase the required optomechanical cooperativity and scattering rate proportionally to the mechanical noise occupancy. Like many other quantum optomechanical protocols, we require the quantum cooperativity to exceed unity~\cite{aspelmeyer_cavity_2014,meenehan_pulsed_2015, maccabe_nano-acoustic_2020,hofer_quantum_2011}. This requirement limits the repetition rate of today's piezo-optomechanical transducers.

\section{Simulation of Bell-state preparation}
\label{sec:simulations}
\begin{figure}
    \centering
    \includegraphics[width = 0.9\linewidth]{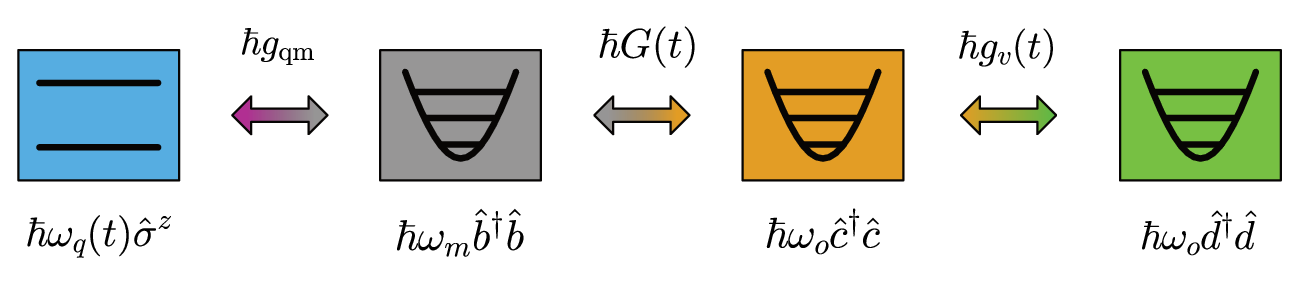}
    \caption{The modes in the QuTiP simulation and their interaction. From left to right we have the transmon qubit, the mechanical mode in the transducer, the optical cavity in the transducer, and a virtual cavity to capture the emission from the optical cavity. The transmon qubit is treated as an anharmonic oscillator when simulated, to include contributions from the higher-excitation states.}
    \label{fig:sim_modes}
\end{figure}

We simulate the evolution of the state of the piezo-optomechanical transducer during microwave-optical entangled-state preparation using a Lindblad master equation \cite{gardiner_input_1985}. The superconducting qubit is treated as an anharmonic oscillator rather than a two-level system when we tune its frequency in and out of resonance with the mechanics. To capture downconverted optical photons leaving the optical cavity inside the transducer, we use the formalism of Ref.~\cite{kiilerich_input-output_2019}, in which a virtual cavity interacts with the optical cavity. By choosing an appropriate coupling parameter $g_v(t)$ between the optical cavity and the virtual cavity, photons from the optical cavity will be absorbed by the virtual cavity and not leak back into the optical cavity~\cite{kiilerich_input-output_2019}. The virtual cavity acts as the output mode from the transducer with which the mechanical mode is entangled. The mode structure and the coupling between the modes are shown in \figref{fig:sim_modes}. 

As a consequence of the fast response of the optical cavity compared to the optomechanical squeezing interaction ($\kappa \gg G(t)$), the output wave packet from the optical mode is well approximated by the shape of the pump pulse. To absorb the output from the optical cavity, the complex coupling between the virtual cavity and the optical cavity is 
\begin{equation}
    g_v(t) = -\frac{v^*(t)}{\int_0^t dt' \vert v(t')\vert^2},
\end{equation}
where
\begin{equation}
    v(t) \propto \exp\left[-\frac{(t-t_\mu)^2}{4\tau_{\mathrm{pulse}}^2}\right]\exp(-i\omega_m t)
\end{equation}
is approximately the shape of the output wave packet from the optical cavity in the rotating frame of the laser pump. Here, $t_\mu$ is the center of the laser pulse and $\tau_{\mathrm{pulse}}$ is the standard deviation of the pulse. 

The master equation takes the form
\begin{equation}
    \Dot{\hat{\rho}} (t) = \mathcal{L}(t)\hat{\rho}(t),
    \label{eq:Lindblad}
\end{equation}
where $\hat{\rho}$ is the combined state of the superconducting qubit, the mechanical transducer mode, the optical cavity, and the virtual cavity representing the entangled waveguide mode (\figref{fig:sim_modes}). The master equation can be written in Lindblad form as
\begin{equation}
    \frac{d\hat{\rho}}{dt} = -\frac{i}{\hbar}\mleft[\hat{H} + \hat{H}', \hat{\rho}(t)\mright] + \sum\limits_{i = 0}^n \mathcal{D}\mleft[\hat{L}_i(t)\mright]\hat{\rho}(t),
\end{equation}
where 
\begin{equation}
    \hat{H} = \hbar\omega_q(t) \hat{a}^\dagger \hat{a} - \frac{E_C}{2}\hat{a}^\dagger\hat{a}^\dagger\hat{a}\hat{a} + \hbar \omega_m \mleft(\hat{b}^\dagger \hat{b} - \hat{c}^\dagger \hat{c}\mright)
    + \hbar g_{\mathrm{qm}}\mleft(\hat{a} + \hat{a}^\dagger\mright) \mleft(\hat{b} +\hat{b}^\dagger\mright)
\end{equation}
is the Hamiltonian describing the transducer coupled to a superconducting qubit with charging energy $E_C$, which determines its anharmonicity~\cite{koch_charge_2007}, and
\begin{equation}
    \hat{H}' = \frac{i\hbar}{2}\sqrt{\kappa}g_v^*(t) \hat{c}^\dagger \hat{d} - \mathrm{H.c.}
\end{equation}
describes the coupling between the optical and virtual cavities. The sum over superoperators $\mathcal{D}(\hat{L}_i)\hat{\rho} = \hat{L}_i\hat{\rho}\hat{L}_i^\dagger - 1/2 \mleft\{\hat{L}_i^\dagger \hat{L}_i, \hat{\rho}\mright\}$ includes the dissipator
\begin{equation}
    \hat{L}_0 = \sqrt{\kappa}\hat{c} + g_v^* \hat{d}.
\end{equation}
For an appropriate choice of $g_v(t)$, the optical and virtual cavity modes will evolve as a dark state of this dissipator. We include other jump operators $\hat{L}_{i}$ representing the incoherent interaction of the transducer with its environment. We use the model introduced by Meenehan \textit{et al.}~\cite{meenehan_pulsed_2015} for the time evolution of the mechanical mode population $n$ while the optical pump is turned on:
\begin{equation}
    \langle \Dot{n}\rangle = \mleft(-\gamma + \Gamma_{\mathrm{OM}}\mright)\langle n\rangle + \gamma_b n_b \mleft(1-\delta e^{-\gamma_s t}\mright) + \Gamma_{\mathrm{OM}}.
    \label{eq:phonon_occupancy}
\end{equation}
Here, $\gamma = \gamma_0 + \gamma_b$ is the total decay rate of the mechanical mode with $\gamma_0$ the coupling to the millikelvin fridge bath. The driving term includes the steady-state occupation $n_b$ of the laser-induced thermal bath, the fraction $\delta$ of the hot thermal bath that is slow to turn on, and the turn-on rate $\gamma_s$ of the hot bath. When the optical pump is turned off, the transducer will cool down towards the ground state at a rate $\gamma_0$ \cite{meenehan_pulsed_2015}. 

\subsection{State-of-the-art transducer design}
\label{sec:SOTA_transducer}
In this section, we benchmark the performance of microwave-optical Bell-state production in our scheme using the transducer design from Ref.~\cite{chiappina_design_2023}. Later we will give an example of a transducer that will allow microwave-optical Bell-state preparation close to the threshold for fault-tolerant quantum computation and with very high ($\sim\SI{20}{\percent}$) heralding probability. We do not assume that we have access to superconducting qubits that preserve their coherence in the presence of an optical pulse, but we assume that the qubits in our architecture have recovery times on the order of \SI{10}{\nano\second}, or that the qubits are shielded from the stray optical photons coming from the pump. Achieving proper shielding and fast recovery of the superconducting qubit under optical illumination is an active research area where significant progress is being made by using gap engineering and low-quasiparticle-lifetime materials. An overview of the performance parameters of state-of-the-art transducers used in our simulation is shown in Table I in the main text.

We use \eqref{eq:phonon_occupancy} to estimate the expected number of thermal phonons in the mechanical mode after the pump pulse. The transducer cannot be squeezed very hard because of the strong external coupling to the optical cavity. This limits the number of downconverted phonons, and so $\langle n\rangle \lesssim 0.05$. We find that, during the pump pulse, the peak optomechanical scattering rate is \SI{540}{\kilo\hertz}, and so the first term on the right-hand side of \eqref{eq:phonon_occupancy} is $\lesssim \SI{27}{\kilo\hertz}$. The second term is much larger, since $\gamma_b n_b \sim 2\pi \times \SI{100}{\kilo\hertz}$. The third term is just the optomechanical scattering rate, which we found to be similar to $\gamma_b n_b$. Dropping the first term and integrating \eqref{eq:phonon_occupancy}, we find that for a square pulse of duration \SI{50}{\nano\second} (similar in length to the Gaussian pulse in the simulation), the number of phonons should be $\sim 0.03$. This is consistent with the estimate of 0.5 added noise phonons for a pulse of \SI{500}{\nano\second} from Ref.~\cite{chiappina_design_2023}. The number of phonons in the mechanical mode originating from the spontaneous down-conversion process is here expected to be similar to the number of thermal phonons. 

\begin{figure}
    \centering
    \includegraphics[width = 0.8\linewidth]{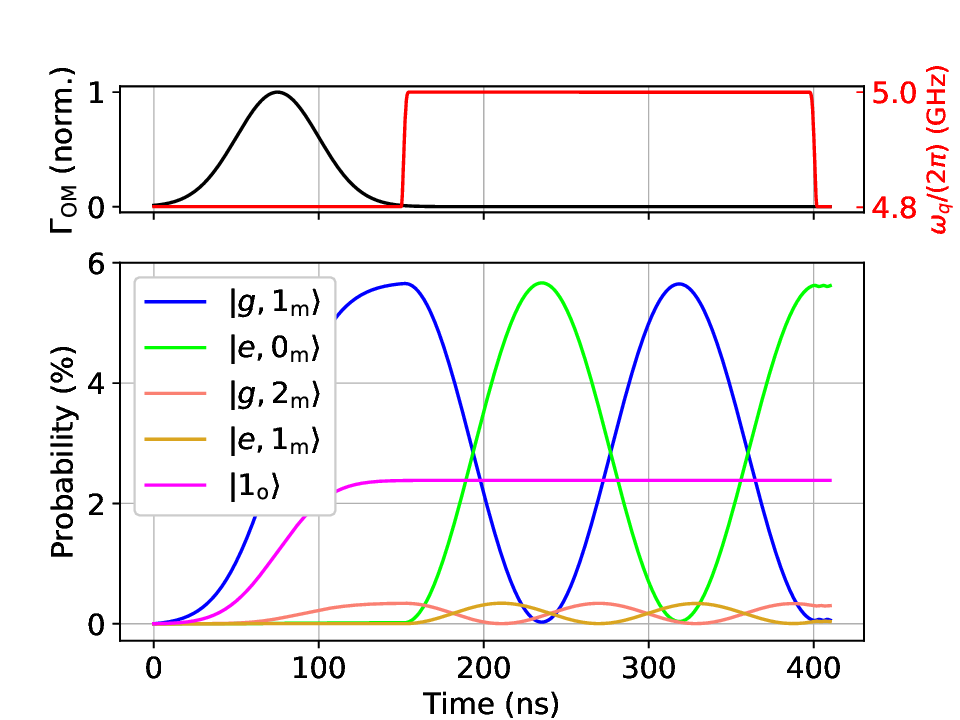}
    \caption{Evolution of the probabilities of finding the transducer system in the state $ \vert n_q n_m\rangle$ using values from Table I in the main text. The probability of finding an optical photon in the virtual cavity representing the output waveguide is also shown.}
    \label{fig:pop_SOTA}
\end{figure}

We now simulate the transducer described above using the QuTiP package~\cite{johansson_qutip_2013} with each mode in \figref{fig:sim_modes} having a four-dimensional Hilbert space. The probability of finding the superconducting qubit and the mechanical mode in the state $\vert n_q n_m\rangle$, where $0_q = g$ and $1_q = e$, is 
\begin{equation}
    P_{n_q n_m}(t) = \langle n_q n_m \vert \hat{\rho}_{\mathrm{qm}}(t) \vert n_q n_m\rangle,
    \label{eq:P}
\end{equation}
where $\hat{\rho}_{\mathrm{qm}}(t) = \mathrm{Tr}_{c,d}\mleft[\hat{\rho}(t)\mright]$ is the state of the superconducting qubit and the mechanical mode after tracing out the optical and virtual cavity modes. Starting from the vacuum state $\hat{\rho}_{\mathrm{qm}} (0) = \vert \mathrm{vac}\rangle\langle \mathrm{vac} \vert$, the evolution of $P_{n_q n_m}$ under \eqref{eq:Lindblad} is shown in \figref{fig:pop_SOTA}. We have included only the relevant states for our scheme; vacuum is the dominant contribution to $\hat{\rho}_{\mathrm{qm}}$, but does not play a role in our scheme, while states with more than two excitations have vanishing contributions. We recognize the expected behavior of the mechanical mode during squeezing, where $P_{02} \approx (P_{01}) ^2$. We have also included the probability of finding one photon in the virtual cavity mode, which represents the optical output from the transducer. 
The simulated probabilities validate our estimate of the thermal phonon contribution to the final transmon population, although the heating appears to be around the upper end of the estimate. This can be attributed mainly to the use of a Gaussian pulse shape in the simulation, which implies that the optical pump is heating the transducer for a longer period than in our crude estimate with a square pulse. We see from \figref{fig:pop_SOTA} that the probability of finding an optical photon in a wave packet emitted from the transducer is roughly \SI{40}{\percent}. We know that $\kappa_{\mathrm{int}}/(\kappa_{\mathrm{ex}}+\kappa_{\mathrm{int}}) = \SI{29}{\percent}$ of optical photons are lost to intrinsic loss channels in the transducer. The remaining observed loss is mainly due to thermal phonons populating the transducer's mechanical mode while the laser is switched on. Heating of the transmon qubit from the microwave photon bath is negligible on the timescale of the simulation. 

\begin{figure}
    \centering
    \includegraphics[width = 0.8\linewidth]{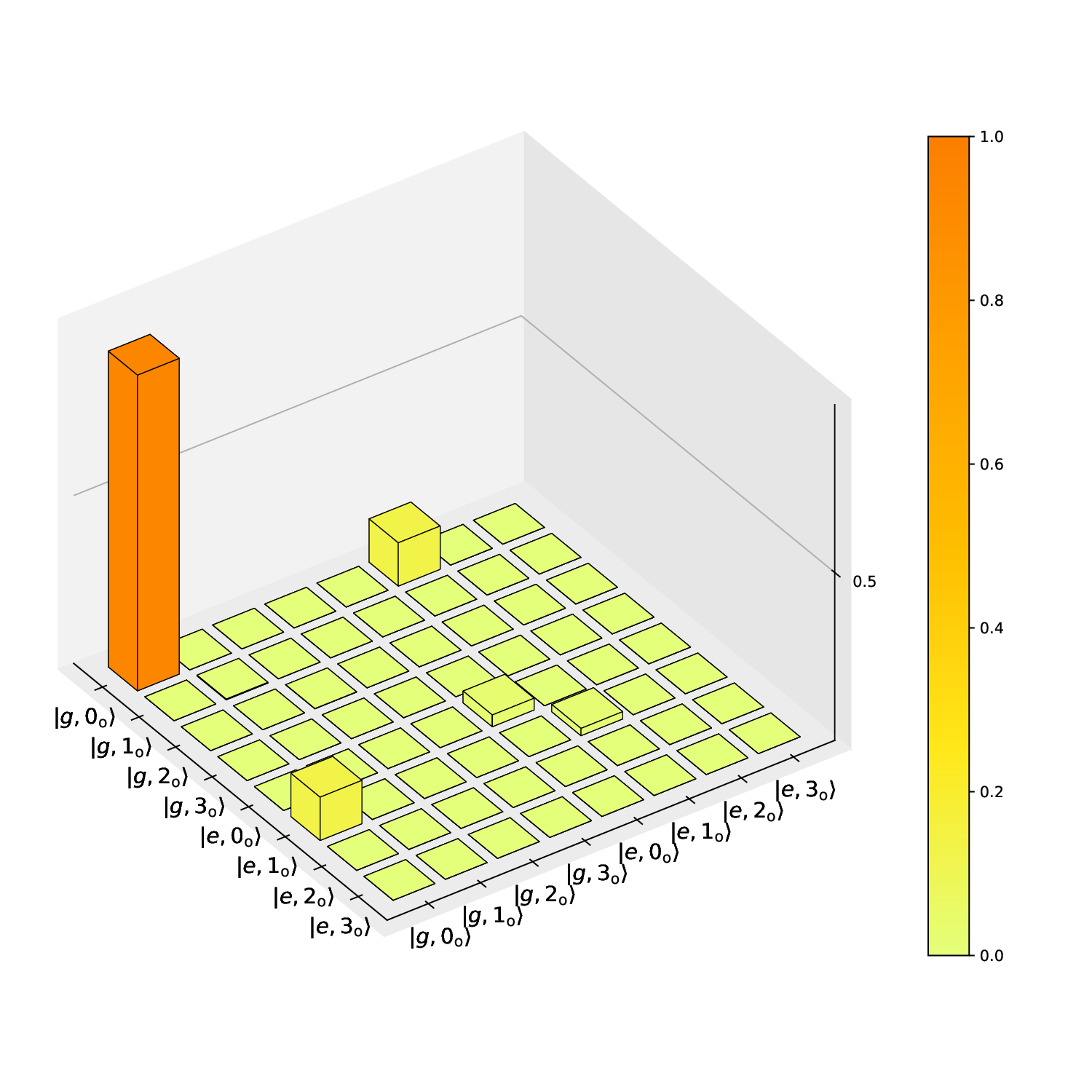}
    \caption{Reduced density matrix of the transmon qubit and the optical output from the transducer at the end of the simulation.}
    \label{fig:matrix_SOTA}
\end{figure}

We are also interested in the coherence of the microwave-optical state which will be used to form one rail in a microwave-optical dual-rail Bell pair. The reduced density matrix of the transmon qubit and the virtual cavity representing the outgoing optical wave packet is shown in \figref{fig:matrix_SOTA}. 
To estimate the fidelity of a heralded Bell pair, we take two copies of the reduced density matrix in \figref{fig:matrix_SOTA} and form a density matrix $\hat{\rho}_{\mathrm{DR}} = \hat{\rho}_{\mathrm{R1}} \otimes \hat{\rho}_{\mathrm{R2}}$. The heralding projects this state onto the subspace where either of the two transmons is excited. Specifically, we take the heralding operation to measure the observable associated with the operator $\hat{Z}_{\mathrm{q1}}\hat{Z}_{\mathrm{q2}}$~\cite{riste_deterministic_2013}. We express the eigenspace of this operator in a tensor product with the identity operator applied to the optical output from the transducer using a basis that includes the target dual-rail microwave-optical state $\vert +\rangle_L = \left(\vert 0_{\mathrm{e}}\rangle\vert 0_{\mathrm{o}}\rangle + \vert 1_{\mathrm{e}}\rangle\vert 1_{\mathrm{o}}\rangle\right)/\sqrt{2}$. This allows us to express the post-heralding density matrix as
\begin{equation}
\begin{split}
    \hat{\rho} &= \frac{1}{P(n_{\mathrm{q, R1}}+n_{\mathrm{q, R2}} = 1)} \Bigg[ \Bigg. P_{\vert +\rangle_L} \vert +\rangle_L\langle + \vert \\
    &+ \frac{1}{2}\sum_{n \neq 1}P_{n,+} \left(\vert 0_{\mathrm{q1}} 0_{\mathrm{o1}} 1_{\mathrm{q2}} n_{\mathrm{o2}}\rangle + \vert 1_{\mathrm{q1}}n_{\mathrm{o1}}0_{\mathrm{q2}} 0_{\mathrm{o2}}\rangle\right) \left(\langle 0_{\mathrm{q1}} 0_{\mathrm{o1}} 1_{\mathrm{q2}} n_{\mathrm{o2}}\vert + \langle 1_{\mathrm{q1}}n_{\mathrm{o1}}0_{\mathrm{q2}} 0_{\mathrm{o2}}\vert\right)\\
    &+ \frac{1}{2}\sum_{n \neq 1} P_{n,-} \left(\vert 0_{\mathrm{q1}} 0_{\mathrm{o1}} 1_{\mathrm{q2}} n_{\mathrm{o2}}\rangle - \vert 1_{\mathrm{q1}}n_{\mathrm{o1}}0_{\mathrm{q2}} 0_{\mathrm{o2}}\rangle\right) \left(\langle 0_{\mathrm{q1}} 0_{\mathrm{o1}} 1_{\mathrm{q2}} n_{\mathrm{o2}}\vert - \langle 1_{\mathrm{q1}}n_{\mathrm{o1}}0_{\mathrm{q2}} 0_{\mathrm{o2}}\vert\right)\\
    &+\dots
    \Bigg. \Bigg]. 
\end{split}
\label{eq:rho}
\end{equation}
The first and second sum in \eqref{eq:rho} capture cases where the transmon qubit is excited but the number of photons in the corresponding optical output mode is more than one or zero---representing multiphoton noise and photon loss, respectively. Here, $P(n_{\mathrm{q, R1}}+n_{\mathrm{q, R2}} = 1)$ is the total probability that a perfect measurement of $\hat{Z}_{\mathrm{q, R1}}\hat{Z}_{\mathrm{q, R2}}$ returns $-1$. The probability of finding a microwave-optical dual-rail Bell pair $\vert +\rangle_L$ is shown in Table~\ref{tab:overlaps_SOTA} along with the most likely other states. 
\begin{table}[b]
    \caption{\label{tab:overlaps_SOTA}
    Overlaps of the simulated state produced from a single block in our scheme with selected candidate states before and after heralding. Before heralding the overlap with the vacuum state is dominant, as can be seen in \figref{fig:matrix_SOTA}.}
    \begin{ruledtabular}
    \begin{tabular}{lccc}
    State & Before heralding & After heralding & Error observed\\
    \hline
        $\vert +\rangle_L$ & $0.041$ & $0.380$ & N/A\\
        $\left(\vert 0_{\mathrm{q1}} 0_{\mathrm{o1}} 1_{\mathrm{q2}} 0_{\mathrm{o2}}\rangle \pm \vert 1_{\mathrm{q1}}0_{\mathrm{o1}}0_{\mathrm{q2}} 0_{\mathrm{o2}}\rangle\right)/\sqrt{2}$ & $0.066$ & $0.614$ & Photon loss\\
        $\left(\vert 0_{\mathrm{q1}} 0_{\mathrm{o1}} 1_{\mathrm{q2}} 2_{\mathrm{o2}}\rangle \pm \vert 1_{\mathrm{q1}}2_{\mathrm{o1}}0_{\mathrm{q2}} 0_{\mathrm{o2}}\rangle\right)/\sqrt{2}$ & $<0.001$ & $0.003$ & Multiphoton noise\\
    \end{tabular}
    \end{ruledtabular}
\end{table}
The poor fidelity of \SI{38}{\percent} is mainly due to the high coupling of the optical cavity in the transducer to intrinsic loss channels and thermal noise. Ignoring finite phonon-photon swap efficiency and multiphoton noise for a moment, the `signal-to-noise' (SNR) ratio in the heralding process is approximately given by $\Gamma_{\mathrm{OM}}\tau_{\mathrm{pulse}}\eta_o/n_{\mathrm{th}}$, where $\eta_o = \kappa_{\mathrm{ex}}/\kappa$ is the optical photon extraction efficiency and $n_{\mathrm{th}}$ is the number of thermal phonons in the mechanical mode of the transducer. Since $\Gamma_{\mathrm{OM}} \propto \kappa^{-1}$, we see that SNR $\propto \kappa^{-2}$. An improvement in the optical quality factor of the transducer would therefore lead to markedly better fidelity of the microwave-optical Bell pair. Indeed, optical quality factors in piezo-optomechanical transducers is the most important area of improvement for practical use of our scheme. Remarkably, the theoretical threshold for photon loss in fault-tolerant linear optical quantum computation is \SI{50}{\percent}~\cite{varnava_loss_2006}. The simulated device is thus closer to being compatible with fault-tolerant quantum computation than what might be believed from the low fidelity alone---although the overhead required would be completely impractical with photon loss rates around \SI{50}{\percent}. Of course, there are several other challenges that are not captured by the simulation here, including high-fidelity heralding, microwave processing after heralding and distinguishable optical photons coming from different devices. 

\subsection{Improved transducers}
\label{sec:improved_transducer}
In the main text, we argue that our scheme can produce resource states from microwave-optical Bell pairs at the threshold for fault-tolerant fusion- or measurement-based quantum computation. Here, we simulate the preparation of a microwave-optical Bell state using transducers that have better performance on three key parameters: intrinsic optical quality factor, mechanical mode heating, and electromechanical coupling. Although the precise mechanism for mechanical mode heating in optomechanical crystals is still under investigation, there is reason to believe that higher intrinsic optical quality factors will reduce the sources of mechanical heating as well. Moreover, a quasi-2D optomechanical crystal with hot phonon bath population $n_b$ reduced by more than a factor of 7 compared to the conventional nanobeam design was recently demonstrated~\cite{ren_two-dimensional_2020}. We do not want to understate the challenge in making transducers with this kind of performance, but such demonstrations indicate that a fundamental redesign of transducers in order to match the parameters used in our simulations is not necessary.

\begin{figure}
    \centering
    \includegraphics[width=\linewidth]{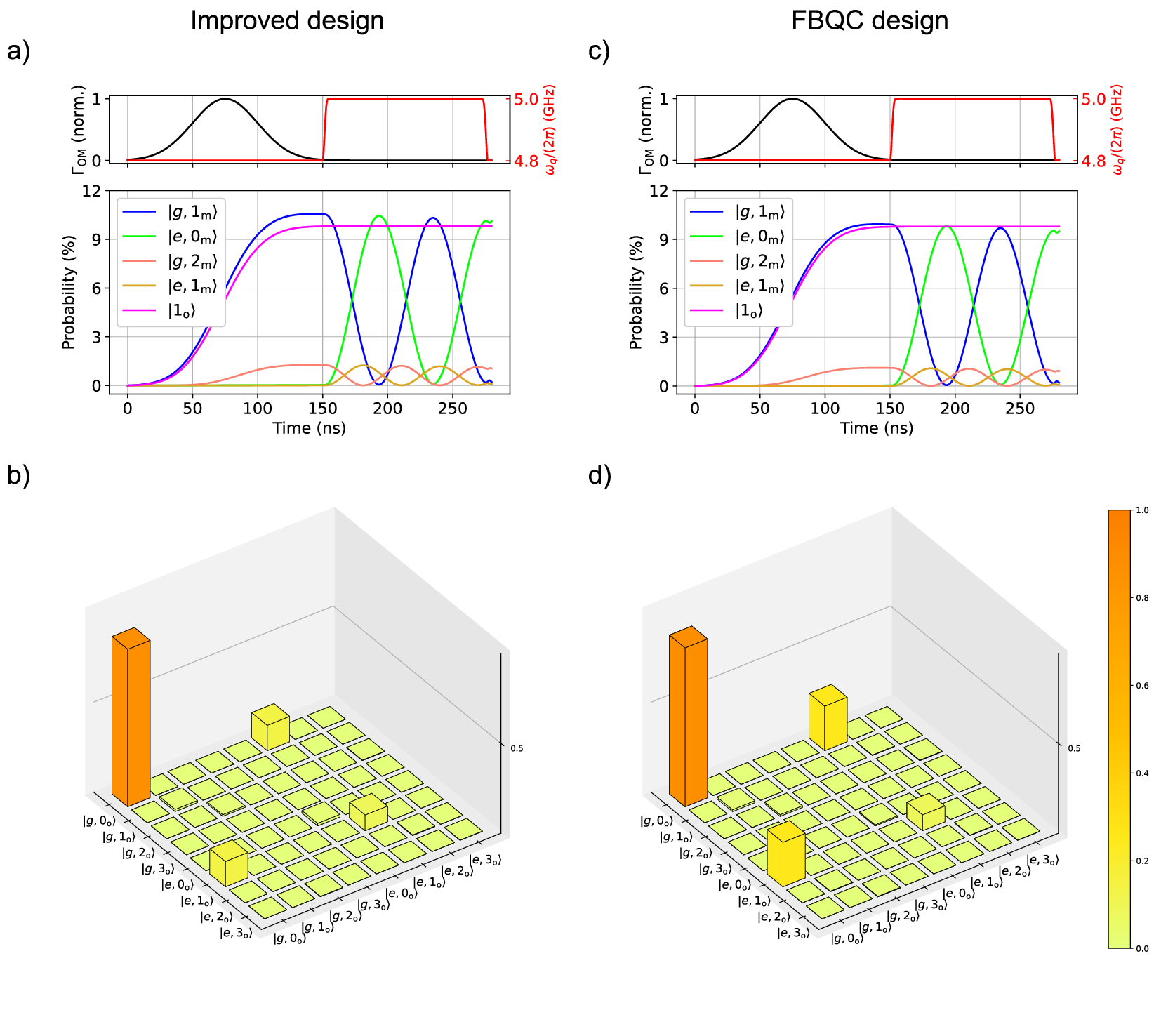}
    \caption{Evolution of the transducer mode populations (a, c) during two-mode squeezing and subsquent phonon swap to the transmon qubit, using using parameters from Table II of the main text. Density matrices (b, d) for one rail after the swap operation from the transducer to the transmon qubit.}
    \label{fig:pop_good}
\end{figure}

We rerun the simulation in Section \ref{sec:SOTA_transducer} with updated parameters for transducer and qubit performance as shown in Table I of the main text.
In addition to decreasing the intrinsic loss of the optical cavity, we have also increased the mechanical mode's loss to the fridge bath by a factor 10. Ultrahigh mechanical quality factors, such as those demonstrated in optomechanical crystals~\cite{maccabe_nano-acoustic_2020, ren_two-dimensional_2020} are not necessary for our scheme provided the electromechanical coupling is sufficiently strong. Low mechanical loss rates reduce the repetition rate of our scheme, which is undesirable for resource-intensive, long-term applications such as quantum communication and (especially) quantum computation. The simulated populations of the qubit and transducer modes, as well as the density matrices at the end of each simulation, are shown in \figref{fig:pop_good}.   
Each mode has been simulated with a four-dimensional Hilbert space. The squeezing is so strong that it introduces significant components of the two- and three-photon states that can lead to multiphoton noise. Four-photon contributions are not captured by the simulation but will only marginally alter the results. Increasing the dimensions of the Hilbert space of the modes is computationally demanding and the additional accuracy does not reflect the uncertainty in the parameters for the transducer in the simulation. Comparing this result with \figref{fig:pop_SOTA}, we see that a larger share of the total contribution to the density matrix comes from the state $\vert e, 1_{\mathrm{o}}\rangle$. This increase in SNR confirms the importance of reducing the intrinsic optical loss of the transducer. However, the strong squeezing leads to undesirable populations along the diagonal on the density matrix, which in turn lead to multiphoton noise in the heralded microwave-optical Bell pair. 
The simulation shows that the majority of heralding events correspond to a true microwave-optical Bell pair. The dominant error source is optical photon loss, either from thermal phonons being swapped to the qubit, or from scattered photons being lost to one of the intrinsic loss channels in the optical cavity of the transducer. However, the strong squeezing is producing a significant amount of multiphoton noise as well. The swap time has not been fully optimized to reduce multiphoton noise in this simulation.

\section{Error processes}
In this section, we present a more detailed description of error processes in the hardware and their effects on resource states. Analyses and measurements of error rates of dual-rail qubits and cavities have been performed elsewhere~\cite{kubica_erasure_2023,Levine2024,teoh_dual-rail_2023,chou_demonstrating_2023}. Many of these results apply to the superconducting architecture of our scheme. The scheme could be adapted to cavities instead of transmon qubits forming the rails. However, the requirements for microwave dual-rail qubits in our scheme are less demanding than for dual-rail qubits used in direct processing of quantum information, since leakage in the microwave dual-rail qubit will be detected when we measure the state of the qubits. Should our measurements indicate that leakage has occurred, we may erase all affected qubits before they are used in a computation on a photonic quantum processor. Errors in the heralding process could lead to multiphoton noise in the optical qubits, but this will not lead to logical errors in the computation. Such leakage will be detected when the optical qubits are fused or measured, as long as the error rate is below thresholds for fault tolerance in fusion- or measuremement-based computational schemes~\cite{bartolucci_fusion-based_2023, briegel_measurement-based_2009, bell_optimizing_2023}.

In the main text, we explain the first-order error mechanisms in our scheme and their effect on the resource states. We focus primarily on the matter-based parts of our scheme (i.e., the mechanical mode and the microwave dual-rail qubit) because we expect the optical side to be dominated by photon loss. The optical photons will not be completely indistinguishable due to disorder which leads to errors when the photons are interfered. Such disorder in the optical and mechanical frequencies may be addressed through, e.g., frequency shifting \cite{riedinger_remote_2018} or tuning.

We summarize the expected error sources and their contribution to errors in the optical output of the resource states in Table~\ref{tab:errors}. 
\begin{table}[b]
\caption{\label{tab:errors}%
    Estimated error sources and effect on optical qubits in the resource states. The values given are order-of-magnitude estimates after heralding and will depend on architecture, hardware performance, and the speed of qubit gates and measurements. See the simulation results presented in Section~\ref{sec:simulations} for two concrete examples covering just the Bell state preparation and heralding. We discuss only first-order effects; chains of errors that produce erasure could combine into Pauli errors, but such chains individually are unlikely to occur. Here, $t$ is the execution time of one clock cycle, $t_{\mathrm{swap}}$ is the swap time from mechanical mode to superconducting qubit, $\gamma$ is the total mechanical decay rate, $T_{\phi, \mathrm{m}}$ is the effective mechanical dephasing rate, $T_{\phi, \mathrm{DR}}$ is the dual-rail qubit dephasing time, $T_{1, \mathrm{DR}}$ is the dual-rail qubit $T_1$, $\eta_{\mathrm{PC}}$ is the parity-check fidelity, and $\eta_{\mathrm{RO}}$ is the readout fidelity for a single superconducting qubit. All other parameters are as defined before.}
\begin{ruledtabular}
\begin{tabular}{lcccc}
Error source & Effect & Scaling & Value & Error type\\ 
\hline
Photon extraction & Photon loss & $\kappa_{\mathrm{int}}/(\kappa_{\mathrm{ex}} + \kappa_{\mathrm{int}})$ & 10\% & Erasure\\
Thermal noise & False heralding & $n_b\gamma_b \left(\tau + t_{\mathrm{swap}}\right)$ & 10\% & Erasure\\
Hard squeezing & Multiphoton noise & $\left(\Gamma_{\mathrm{OM}}\tau\right)^2$ & 1\% & Leakage\\
Phonon loss & Multiphoton noise & $\gamma\left(\tau + t_{\mathrm{swap}}\right)$ & 1\% & Leakage\\
Imperfect parity check & False heralding & $1-\eta_{\mathrm{PC}}$ & 1\% & Erasure\\
Measurement infidelity & Inconsistent measurement & $1-\eta_{\mathrm{RO}}$ & 1\% & Erasure\\
Mechanical dephasing & Phase flip & $\left(\tau + t_{\mathrm{swap}}\right)/T_{\phi,\mathrm{m}}$ & 1\% & Pauli error\\
Qubit dephasing & Phase flip & $t/T_{\phi, \mathrm{DR}}$ & 0.1\% & Pauli error\\
Qubit swap & Bit flip & $t/T_{1, \mathrm{DR}}$ & 0.1\% & Pauli error\\
Measurement infidelity & Phase/bit flip & $\left(1-\eta_{\mathrm{RO}}\right)^2$ & 0.01\% & Pauli error\\

\end{tabular}
\end{ruledtabular}
\end{table}
As discussed in the main text, we expect that a low extraction efficiency from the optical cavities will result in photon loss rates on the order of \SI{10}{\percent}. Pump photons will have to be filtered out from the optical output, which will further increase the photon loss rate. Further, for the mechanical mode and the superconducting qubits, there are several sources of errors with varying degrees of severity. Thermal noise in the mechanical mode will primarily contribute to false heralding of a Bell pair, which is equivalent to a correctly heralded Bell pair with its optical photon missing. Thermal phonons may appear in the mechanical mode while the Bell pair is being prepared, or the mechanical mode might not be fully in its ground state at the beginning of a clock cycle. The latter could be the dominant contribution, particularly when operating the resource-state generator at high rates.  

When we pump the optomechanical crystal to produce phonon-photon pairs, there is a risk of producing more than one such pair from the same pump pulse. As is evident from the simulations presented in Section~\ref{sec:simulations}, this will lead to multiphoton noise in the optical qubits. We propose to suppress this noise by using the mechanics-qubit swap to avoid heralding from two-phonon states. While the swap time can be optimized to suppress noise, it is challenging to eliminate noise from both two-phonon and three-phonon states simultaneously. Assuming the optimal strategy would be to eliminate noise from two-phonon states because these are much more likely to be produced, we can estimate the level of three-photon noise from three downconverted pairs. For simplicity we take the optomechanical crystal after the pump pulse to be in a perfect two-mode squeezed vacuum state,
\begin{equation}
    \vert \psi\rangle = \sqrt{p_0}\vert 0_m0_o\rangle + \sqrt{p_1}\vert 1_m1_o\rangle +\sqrt{p_2}\vert 2_m2_o\rangle + \sqrt{p_3}\vert 3_m3_o\rangle + \cdots,
\end{equation}
where $p_n$ is the probability of finding $n$ phonon-photon pairs. We neglect states with four or more such pairs. From Eq.~(3) in the main text, it follows that $p_n = p_0(1-p_0)^n$. Next, we count the different combinations of states that will herald a Bell pair. Our heralding scheme cannot distinguish one-phonon states from three-phonon states, nor can it distinguish two-phonon states from zero-phonon states. Combining the optomechanical transducers from both rails, we have
\begin{multline}
    \vert \psi_{\mathrm{DR}}\rangle = p_0 \vert 0_{\mathrm{R1}}0_{\mathrm{R2}}\rangle + \sqrt{p_0p_1}\vert 1_{\mathrm{R1}}0_{\mathrm{R2}}\rangle + \sqrt{p_0p_1}\vert 0_{\mathrm{R1}}1_{\mathrm{R2}}\rangle + \cdots \\
    + \sqrt{p_0p_3}\vert 0_{\mathrm{R1}}3_{\mathrm{R2}}\rangle + \sqrt{p_0p_3}\vert 3_{\mathrm{R1}}0_{\mathrm{R2}}\rangle + \sqrt{p_1p_2}\vert 1_{\mathrm{R1}}2_{\mathrm{R2}}\rangle+ \sqrt{p_1p_2}\vert 2_{\mathrm{R1}}1_{\mathrm{R2}}\rangle + \cdots.
    \label{eq:two_rails}
\end{multline}
Here, $\vert n_{\mathrm{R1}}m_{\mathrm{R2}}\rangle$ is the state with $n$ phonon-photon pairs in rail 1 and $m$ phonon-photon pairs in rail 2. We only write explicitly the terms with $n+m = 0, 1, 3$. States with $n + m = 2$ are taken care of by the heralding, while states with $ n + m > 3$ are ignored. Since all the terms in \eqref{eq:two_rails} apart from $\vert 0_{\mathrm{R1}}0_{\mathrm{R2}}\rangle$ will herald a Bell pair, we find that the probability of having three photons in the optical rails is
\begin{equation}
    P_{\mathrm{noise}} = \frac{p_0p_3 + p_1p_2}{p_0p_1 + p_0p_3 + p_1p_2} \approx \left(1-p_0\right)^2 \approx p_1^2.
\end{equation}
Following Ref.~\cite{meenehan_pulsed_2015}, we can estimate how $p_1$ scales with $\Gamma_{\mathrm{OM}}\tau$ by defining the propagator $U = \exp\mleft[ir\mleft(\hat{b}(0)\hat{C}_{\mathrm{in}} + \hat{b}^\dagger(0)\hat{C}_{\mathrm{in}}^\dagger \mright)\mright]$. Using
\begin{equation}
    p_1 \approx \mleft\vert \langle 1_b1_C\vert U\vert 0_b0_C\rangle\mright\vert^2 \sim r^2,
\end{equation}
we find from expanding both sides of $\cosh(r) = e^{\Gamma_{\mathrm{OM}}\tau/2}$ to lowest order that
\begin{equation}
    p_1 \sim r^2 \sim \Gamma_{\mathrm{OM}}\tau.
\end{equation}
Note that this holds only for low levels of squeezing. For stronger squeezing ($p_1 > 10\%$) there will be deviations from this lowest-order result. 

Phonon loss can also contribute to multiphoton noise in the optical qubits by converting a two-phonon state into a one-phonon state, which will herald a Bell pair in the parity check. This is technically a second-order effect, because $p_2$ is small compared to $p_1$ and phonons should be quickly swapped to the superconducting qubits, but for hard pumping ($p_1 \sim 10\%$) and lossy mechanics ($\gamma/(2\pi) \sim $ \SI{500}{\kilo\hertz}), this could lead to multiphoton noise and therefore leakage rates of order 1\%.  

Infidelity in the parity check will predominantly mistake an even parity for an odd parity. The probability of such events will depend on how the parity check is implemented, but we assume a false-positive rate of order 1\% is readily achievable using, e.g., the readout architecture in the main text. We are overwhelmingly likely to catch parity-check errors when the dual-rail qubit is measured. If a measurement of the two superconducting qubits forming the microwave dual-rail qubit indicates that a microwave photon is missing (or there is a microwave photon in each superconducting qubit), we erase this dual-rail qubit (and its entangled optical dual-rail qubit) from the resource state. 

A related error source is the measurement of the superconducting qubit in each rail, which will have some fidelity $\eta_{\mathrm{RO}}$. Because we measure each rail independently, the probability of both measurements yielding the wrong answer is $\left(1-\eta_{\mathrm{RO}}\right)^2$, which will lead to a Pauli error. If only one measurement yields the wrong result, then the two measurements will not correspond to a valid dual-rail qubit state and we therefore erase the qubit from the resource state. Qubit readout is an essential part of any superconducting processor, and error rates are already below 1\% in state-of-the-art processors with transmon qubits~\cite{chen_transmon_2023}. 

Among the mechanisms we considered so far, we believe that the primary source of Pauli errors will be dephasing of the mechanical mode before the phonon can be swapped to the superconducting qubit. Investigations into how dephasing times in the mechanical mode can be extended are encouraged. However, the relatively low phonon number in our scheme implies that the mechanical dephasing effect is reduced compared to the single-phonon dephasing rate $\gamma_\phi$. As an illustration, consider a mechanical mode that is subject to frequency fluctuations. In a frame rotating at the mechanical frequency, the mechanical mode can be described by the master equation
\begin{equation}
    \frac{d\hat{\rho}}{dt} = \mathcal{D}\mleft[\sqrt{\gamma_\phi}\hat{b}^\dagger\hat{b}\mright]\hat{\rho}. 
    \label{eq:mechanicsME}
\end{equation}
Suppose the resonator at time $t_0$ is in the pure state $\hat{\rho} = \mleft(\sqrt{p_0}\vert0\rangle + \sqrt{p_1}\vert 1\rangle\mright)\mleft(\sqrt{p_0}\langle 0\vert + \sqrt{p_1}\langle 1\vert\mright)$. In this case, \eqref{eq:mechanicsME} becomes
\begin{equation}
    \frac{d\rho}{dt} = -\frac{\gamma_\phi}{2}\sqrt{p_0p_1} \mleft(\vert 0\rangle\langle 1\vert + \vert 1\rangle\langle 0\vert\mright).
\end{equation}
This describes exponential decay of the off-diagonal terms in the density matrix at a rate that depends on the excited-state population $p_1$. In our protocol, $p_0p_1 \leq 0.1$ and therefore the effective dephasing rate is smaller by at least a factor two compared to cases where the mechanical oscillator is used as a (single-rail) qubit. Higher levels of optomechanical squeezing cause more rapid dephasing, and there might be additional, second-order processes that can lead to dephasing in the strong-squeezing case simulated in Section~\ref{sec:improved_transducer}. For example, we suspect that optical photon loss ($\lesssim \SI{10}{\percent}$) combined with thermal noise and multiphoton noise (which can add up to $\sim\SI{10}{\percent}$) could have a similar contribution to the simulated phase flip rate. 

After the phonon-photon swap operation, there will be dephasing of the microwave dual-rail qubit before it is measured. Promising investigations into such dephasing have begun recently~\cite{Levine2024}. To minimize such errors, we should minimize the time from Bell-state preparation until the microwave dual-rail qubits are measured and maximize $T_{\phi}$ for the dual-rail qubit. The time from start to end of a single clock cycle as described in the main text will be similar to a surface-code cycle~\cite{fowler_surface_2012}, which is likely to be around \SI{1}{\micro\second}. With single-transmon dephasing times up to \SI{100}{\micro\second} \cite{kosen_building_2022,place_new_2021}, dual-rail $T_{\phi}$ of order \SI{10}{\milli\second} should be possible~\cite{kubica_erasure_2023}. We expect that Pauli error from decoherence of the superconducting dual-rail qubit can thus be kept well below the threshold for fault tolerance.

\section{Two-qubit gates between hybrid qubits}
We define hybrid qubits in the main text with a repetition code using the microwave and optical dual-rail qubits. Specifically, our protocol prepares the hybrid states 
\begin{equation}
    \vert +\rangle_H = \frac{1}{\sqrt{2}}\mleft(\vert 0_\mathrm{e}\rangle\vert 0_\mathrm{o}\rangle + \vert 1_\mathrm{e}\rangle\vert 1_\mathrm{o}\rangle\mright),
\end{equation}
where the dual-rail qubits are encoded as ($i = \mathrm{e,o}$) 
\begin{eqnarray}
    \vert 0_i\rangle &=& \frac{1}{\sqrt{2}}\mleft(\vert 0_i\rangle_{\mathrm{R1}}\vert 1_i\rangle_{\mathrm{R2}} + \vert 1_i\rangle_{\mathrm{R1}}\vert 0_i\rangle_{\mathrm{R2}}\mright), \\
    \vert 1_i\rangle &=& \frac{1}{\sqrt{2}}\mleft(\vert 0_i\rangle_{\mathrm{R1}}\vert 1_i\rangle_{\mathrm{R2}} - \vert 1_i\rangle_{\mathrm{R1}}\vert 0_i\rangle_{\mathrm{R2}}\mright),
\end{eqnarray}
when expressed as the Fock state of the rails R1 and R2 in the microwave (e) and optical (o) domain. Building graph states requires that we perform CZ gates between microwave qubits belonging to different logical states $\vert +\rangle_H$. This can be done in a straightforward way by bringing the state $\vert 1_\mathrm{e}\rangle\vert 1_\mathrm{e}\rangle$ into resonance with a non-computational state~\cite{dicarlo_demonstration_2009}. While the phase between two dual-rail qubits could be affected while idling at the avoided crossing, this can be corrected with single-qubit gates~\cite{campbell_universal_2020}. We expect that CZ gates performed this way may achieve better fidelities than transmon-transmon CZ gates due to the dual-rail qubits' protection from dephasing~\cite{campbell_universal_2020,Levine2024}. 

Alternatively, we may implement CZ gates from $\sqrt{i\mathrm{SWAP}}$ gates between dual-rail qubits and single-qubit gates. For example, we can adiabatically turn on the interaction $g_c$ between two transmons from different blocks to realize the effective Hamiltonian~\cite{kubica_erasure_2023}
\begin{equation}
    \hat{H}_{\mathrm{eff}} = \frac{\Omega}{2}\mleft(\hat{\sigma}^z_1 + \hat{\sigma}^z_2\mright) + g_{XX}\hat{\sigma}^x_1\hat{\sigma}^x_2,  
\end{equation}
where $\Omega = 2g_{12}\mleft(1+6g_c^2/\Delta^2\mright)$ and $g_{XX} = -4g_c^2E_C/\mleft(\Delta^2-E_C^2\mright)$ when the transmons within each block are on resonance and coupled at rate $g_c$, but transmons from different blocks are detuned by $\Delta$. The Pauli matrices $\hat{\sigma}^z_i, \hat{\sigma}^x_i$ generate rotations of the dual-rail qubits $i = 1,2$ around the z- and x-axis, respectively. After a time $T$ such that $\int_0^T g_{XX}(t) dt= \pi/4$, $\hat{H}_{\mathrm{eff}}$ has implemented the gate $\sqrt{i\mathrm{SWAP}}\exp\left[-i\phi\left(\hat{\sigma}^z_1 + \hat{\sigma}^z_2\right)\right]$, where $\phi = \int_0^T \sqrt{\Omega^2 + g_{XX}^2} dt$~\cite{kubica_erasure_2023}. Applying gates to the microwave dual-rail qubit in each hybrid qubit, the following circuit then implements the CZ gate on two logical qubits:\\

\begin{center}
    \begin{quantikz}
        \lstick{$\ket{+}_H$} & \gate{X} & \gate{R_Y(\frac{\pi}{2})} & \gate[2]{\sqrt{i\textsc{swap}}} & \gate{X}& \gate[2]{\sqrt{i\textsc{swap}}}&\gate{R_X(-\frac{\pi}{2})} & \gate{R_Y(-\frac{\pi}{2})}& \gate{X} & \qw\\
\lstick{$\ket{+}_H$} & \gate{H} & \qw & & \qw & & \gate{R_X(-\frac{\pi}{2})}&\qw&\gate{H} & \qw\\
    \end{quantikz}
\end{center}

\section{Turning on the transmon-transmon coupling}
After the parity check (see main text), we have the state $\vert +\rangle_H = \frac{1}{\sqrt{2}}\left(\vert 0_{\mathrm{e}}\rangle\vert 0_{\mathrm{o}}\rangle + \vert 1_{\mathrm{e}}\rangle\vert 1_{\mathrm{o}}\rangle\right).$ The $\vert 0_{\mathrm{e}}\rangle$ and $\vert 1_{\mathrm{e}}\rangle$ states lose their degeneracy as we turn on the transmon-transmon coupling. In addition, $\vert +\rangle_H$ is not an eigenstate of the Hamiltonian governing the system. An ideal coupler would allow us to realize a Hamiltonian where there are no transitions between the states $\vert 0_\mathrm{e}\rangle$ and $\vert 1_\mathrm{e}\rangle$, nor between any computational and non-computational state, thus allowing us to tune the coupling strength arbitrarily fast (confirmed through a QuTiP simulation~\cite{johansson_qutip_2013}). However, tunable couplers are multilevel systems where unwanted transitions can occur. This implies that calibration is necessary to find an optimal tuning speed that maximizes fidelity~\cite{Yan2018}.

\end{document}